\documentclass[preprint]{aastex}
\begin{document}
\title{A Far Ultraviolet Archival Study of Cataclysmic Variables:
I. {\it{FUSE}} and {\it{HST}}/STIS Spectra of the Exposed White Dwarf 
in Dwarf Nova Systems.  
\altaffilmark{1}}

\author{Patrick Godon\altaffilmark{2}, Edward M. Sion} 
\affil{Department of Astronomy and Astrophysics
Villanova University,
Villanova, PA 19085,
patrick.godon@villanova.edu; edward.sion@villanova.edu}

\author{Paul E. Barrett} 
\affil{United States Naval Observatory, 
Washington, DC 20392
barrett.paul@usno.navy.mil} 

\author{Ivan Hubeny}
\affil{Department of Astronomy,
University of Arizona, AZ,
hubeny@aegis.as.arizona.edu} 

\author{Albert P. Linnell, Paula Szkody}
\affil{Department of Astronomy,
University of Washington,
Seattle, WA 98195,
linnell@astro.washington.edu;
szkody@astro.washington.edu} 

\altaffiltext{1}
{Based on observations made with the 
NASA-CNES-CSA Far Ultraviolet Spectroscopic
Explorer. {\it{FUSE}} is operated for NASA by the Johns Hopkins University under
NASA contract NAS5-32985} 
\altaffiltext{2}
{Visiting at the Space Telescope Science Institute, Baltimore, MD 21218,
godon@stsci.edu}

\begin{abstract}

We present a synthetic spectral analysis of Far Ultraviolet Spectroscopic
Explorer ({\it{FUSE}}) and Hubble Space Telescope/Space Telescope Imaging
Spectrograph ({\it{HST}}/STIS)  
spectra of 5 dwarf novae above and below the period gap during quiescence.
We use our synthetic spectral code, including options for 
the treatment of the hydrogen quasi-molecular
satellite lines (for low temperature stellar atmospheres), 
NLTE approximation (for high temperature stellar atmospheres),  
and for one system (RU Peg) we model the interstellar medium 
(ISM) molecular and atomic hydrogen lines.  
In all the systems presented here the FUV flux continuum is due to the
WD. These spectra also exhibit some broad emission lines. In this work
we confirm some of the previous FUV analysis results but we also present new
results. For 4 systems we combine the {\it{FUSE}} and STIS spectra to cover a
larger wavelength range and to improve the spectral fit.    
This work is part of our broader {\it{HST}} archival research program, in which
we aim to provide accurate system parameters for cataclysmic
variables above and below the period gap by combining {\it{FUSE}} and {\it{HST}}
FUV spectra. 

\end{abstract}

\keywords{Stars: white Dwarfs, Stars: dwarf novae 
(SS Aur, EY Cyg, VW Hyi, RU Peg, EK TrA).}  

\section{Introduction}
\subsection{Dwarf Novae in Cataclysmic Variables} 

Cataclysmic variables (CVs) are short-period, semi-detached compact 
binaries in which the primary star, a white dwarf (WD), accretes 
matter and angular momentum from the secondary,  
a main-sequence star filling its Roche lobe (the mass donor) 
\citep{war95}. The matter is transferred by means of either an accretion
disk around the WD, or an accretion column or curtain - when the WD has a 
strong magnetic field.   
Ongoing accretion at a low rate (quiescence) can be interrupted
every few weeks to months by intense accretion (outburst) of days to
weeks (a dwarf nova accretion event), and every few thousand years by a 
thermonuclear runaway explosion (TNR - the classical nova event, due
to the ignition of the accreted layer of hydrogen-rich material). 
CVs are divided in sub-classes according to the duration, occurrence
and amplitude of their outbursts. The two main types of CVs are
Dwarf Nova systems (DNs; weakly- or non-magnetic disk systems 
found mostly in their quiescent state which lasts
much longer than their outburst), and nova-like (NL) systems
which form a less homogeneous class. 
NLs includes disk systems found mostly in their high state,
intermediate polars (IPs; with a magnetically truncated inner disk), 
polars (devoid of disk due to the strong magnetic field of the WD),
and other systems e.g. that never go into outburst or which cannot
be classified as DNs \citep{war95} (this includes the helium systems -
AM CVn - with the shortest period).  
Dwarf nova systems includes the U Gem systems, the SU UMa systems, 
and the Z Cam systems. 
The U Gem systems are the typical DNs, i.e. those systems exhibiting
normal DN outbursts; 
the SU UMas exhibit both normal DN outbursts, and
superoutbursts, which are both longer in duration and higher in 
luminosity than normal DN outbursts; 
and the Z Cam systems have standstills where they remain in a 
state of intermediate optical brightness
for a long time (here we adopt the classification according to
\citet{rit03}). 

Interestingly enough, while the binary orbital period in CV  
systems ranges from a fraction of an hour (AM CVn systems)  
to about a day (e.g. GK Per), there is a gap in the orbital period
between 2 and 3 hours where almost no systems are found 
(hereafter the ``period gap''). For example U Gem and Z Cam
DN systems are found above the period gap, while the SU UMa DN systems
are found below the period gap.  
It is not known 
whether the systems are evolving from a longer period to a shorter
period (across the gap) or whether the systems above the gap are altogether
different from the systems below the gap.

The now widely accepted interpretation of the quiescence/outburst
cycle is the disk instability model (DIM, \citet{can98}).  
It is assumed that during the quiescent
phase the matter in the disk is cold and neutral and the
disk is optically thin because of its low density, while 
during outburst the matter in the disk is ionized and becomes optically thick 
as the mass accretion rate within the disk increases. 
The basic principle of the DIM theory depends 
heavily on the unknown viscosity parameter $\alpha$ \citep{sha73} and 
on the mass transfer rate during the different phases. 
DNs, unlike other CVs, offer a fairly reliable estimate of
their distances via the absolute magnitude at maximum versus orbital
period relation for     DNs     found by \citet{war95}. This relationship
is consistent with theory \citep{can98}. 
The mass accretion rate within the disk has been taken from \citet{pat84}, 
which is, however, only a first order estimate. 
In the last decade, the disk  
mass accretion rate of many systems has 
been deduced more accurately at given epochs
of outburst or quiescence using spectral fitting
techniques. The accretion rate is usually a function of time
(especially during the outburst itself; and it is also a function
of radius $r$ due to wind-outflow from the disk) and, consequently,  
it is difficult to assess its time-averaged value accurately. 

Recent advances in theory \citep{tow04}  have shown that
the average mass accretion rate onto the WD in DNs can be
deduced if one knows the mass of the accreting WD and
its effective surface temperature during quiescence, therefore
providing an additional and independent way to assess $\dot{M}$ 
(or more precisely $\dot{M}(r=R_*)$ - the mass accretion rate
at one stellar radius $R_*$ 
{\it onto} the stellar surface, which might be different than the
mass accretion rate in the disk $\dot{M}(r)$ at a radius $r$
if there is an outflow from the disk: $\partial \dot{M}/\partial r <0$. ). 
Consequently, in order to put more constraints 
on the theories we need to know the
properties (mainly the temperature and mass of the WD) 
of these systems above as well as below the period gap. 
There is, however, a critical shortage 
in knowledge of the WD properties (effective temperature $T_{wd}$, 
gravity $Log(g)$, projected rotational velocity 
$V_{rot} sin(i)$, chemical abundances, accretion belts?) 
in     DNs     above the period gap. 
Thus, detailed comparisons of accreting WDs above and 
below the gap cannot be made. 

For systems below the gap, with 
orbital periods near the period minimum \citep{war95}, 
the distribution of temperatures 
are centered at $\sim$15,000K with only a narrow range identified at present. 
This distribution appears to manifest the effect of long term 
compressional heating at a time averaged accretion rate of $2\times 
10^{-11} M_{\odot}$yr$^{-1}$  
(Townsley and Bildsten 2002; Sion et al. 2003). 
It appears that WD $T_{eff}$'s for systems above the gap are higher than WD 
temperatures in systems below the gap, due to the systems above the gap 
having larger disks (with higher mass transfer rates) and more massive 
(somewhat earlier-type) secondaries.  
Some disks may remain optically thick even during quiescence 
so that the WDs are heated to a greater extent than systems below the gap. 
It is not yet known whether the WDs in systems above the gap are
rotating more slowly than WDs in systems below the gap where
presumably the CVs are older and with a longer history of angular momentum
transfer via disk accretion. Thus far the only     DNs     above the gap 
whose     WDs      and disks/boundary layers 
have been analyzed with {\it{FUSE}}, {\it{IUE}} and {\it{HST}} have been 
Z Cam \citep{har05}, RX And \citep{sio01,sep02},  
U Gem \citep{sio98,lon99,fro01}, SS Aur \citep{sio04a},   
EY Cyg \citep{sio04b}, RU Peg \citep{sio02,sio04a} 
and WW Cet \citep{god06b};    
a total of 2 Z Cam systems, 4 U Gem systems and the peculiar
DN WW Ceti (although classified as a U Gem, others suggest
it is a Z Cam).  

Under our broader {\it{HST}} archival research we have started to  
secure accurate system parameters ($\dot{M}$, 
$i$, $M_{wd}$, $T_{eff}$, $V_{rot} sin(i)$, chemical abundances,etc..)
for CVs (DNs and NLs) above and below the period gap by fitting 
synthetic spectral models (WDs and accretion disks) 
to combined {\it{FUSE}}+{\it{HST}} spectra from the
MAST archive.  We have identified 25  
CV systems for which the {\it{FUSE}} and {\it{HST}} (STIS, FOS or GHRS)
spectra  match and can be combined. 
As a part of this {\it{HST}} archival research, we 
present in this paper 
the analysis of {\it{HST}}/STIS and ({\it{FUSE}}) archival spectra 
of 5     DNs     during quiescence using synthetic stellar spectra
together with the {\it{FUSE}} spectrum of RU Peg. 

\subsection{Five Dwarf Nova Systems}

The five DNs    are listed in Table 1 with their system 
parameters as follows: 
column (1) Name, (2) CV subtype,
(3) reddening value E(B-V), (4) distance, (5) orbital period in days, 
(6) orbital inclination in degrees, (7) spectral type of the secondary, 
(8) mass of the primary in solar masses, (9) mass of the secondary in 
solar masses, (10) apparent magnitude in outburst, and (11) apparent 
magnitude in quiescence. The references are listed below the table.
Three systems are above the period gap: EY Cyg, SS Aur and RU Peg;
and two   systems are below the period gap: VW Hyi and EK TrA.  

Among these 5 objects, 4 DNs in quiescence are directly  
from our {\it{HST}} archival program with matching {\it{FUSE}} and {\it{HST}}/STIS spectra: 
EY CYG, SS Aur, VW Hyi,  and EK TrA. 
We supplement this study here with the analysis of 
RU Peg's {\it{FUSE}} spectrum. 
For SS Aur, VW Hyi and EK TrA we assess the WD parameters by modeling for 
the first time for these 3 objects the combined {\it{FUSE}}+STIS spectra. 
For RU Peg's {\it{FUSE}} spectrum, and EY Cyg's combined {\it{FUSE}}+STIS spectrum 
we improve the fittings that were carried out previously. 

For all the spectra we present here, we use the latest 
versions of the stellar model atmospheres and synthetic spectra codes
(see section 3); this includes options for
the treatment of the hydrogen molecular satellite lines (for modeling 
cooler WDs), and NLTE atmosphere models (for modeling hotter WDs). 
We also adopt  the extinction values E(B-V) given in \citet{bru94}  
for the 5 objects and we deredden the spectra accordingly.  
For some systems we identify the ISM molecular absorption lines 
in the {\it{FUSE}} spectra; 
this greatly helps improve the fitting of the chemical abundances 
and WD's temperature. RU Peg has stronger ISM absorption lines and, for
this system only, we model the ISM atomic and molecular hydrogen opacities.  
For these 5 systems we obtain the temperature, projected rotational velocity
and chemical abundances (of C, N, S and Si) of the exposed     WD     
atmosphere.
We also give upper and lower limits for the mass of the WD and 
the distance to the system.

\section{The Archival Spectra}

The observations log is presented in Table 2. 
From the {\it{AAVSO}} (American Association of Variable Stars Observers)
data, we found that all the systems were in optical quiescence
at the time of the observation (see the last column of Table 2).
The STIS spectra of EY Cyg and SS Aur
are snapshots (lasting 600-700s) with a lower resolution and
S/N than the STIS spectra of VW Hyi and EK TrA.  With fluxes  
$\le 1 \times 10^{-14}$ergs$~$s$^{-1}$cm$^{-2}$\AA$^{-1}$, 
both EY Cyg and EK TrA are actually weak {\it{FUSE}} sources, and RU Peg
(though with a {\it{FUSE}} flux ten times higher) is under-exposed with 
an exposure time of only 2800s.

\subsection{The {\it{FUSE}} Archival Spectra}
\subsubsection{Processing the {\it{FUSE}} Data}

{\it{FUSE}}'s optical
system consists of four optical telescopes (mirrors), each 
connected to a Rowland spectrograph. The four diffraction
gratings of the four spectrographs produce four independent spectra on
two detectors. Two mirrors and two gratings are coated
with SiC to provide wavelength coverage below 1020 \AA, and the other
two mirrors and gratings are coated with Al and LiF.  The
Al+LiF coating provides about twice the reflectivity of SiC at
wavelengths $>$1050 \AA, and very little reflectivity below 1020 \AA . 
These are known as the SiC1, SiC2, LiF1 and LiF2 channels.  

All the {\it{FUSE}} spectra presented here were obtained 
through the 30"x30" LWRS Large Square Aperture in TIME TAG mode. 
The data were processed with CalFUSE version 3.0.7 \citep{dix07},
which automatically  handles event bursts.
Event bursts are short periods during an exposure when high count 
rates are registered on one or more detectors. The bursts exhibit  
a complex pattern on the detector, their cause, however, is yet unknown 
(it has been confirmed that they are not detector effects). 
The main change from previous versions of CalFUSE is that now the data
are maintained as a photon list (the intermediate data file - IDF) throughout
the pipeline. Bad photons are flagged but not discarded, so the user can 
examine, 
filter, and combine data without re-running the pipeline. A number of design changes
enable the new pipeline to run faster and use less disk space than before. 
Processing time with CalFUSE has decreased by a factor of up to 10. 

To process {\it{FUSE}} data, we follow the same procedure used  
previously for the analysis of other systems (such as WW Ceti, \citet{god06b});
consequently we give only a short account of this procedure.  
The spectral regions covered by the
spectral channels overlap, and these overlap regions are then used to
renormalize the spectra in the SiC1, LiF2, and SiC2 channels to the flux in
the LiF1 channel. We then produced a final spectrum that covers the
full {\it{FUSE}} wavelength range $905-1187$ \AA. The low sensitivity 
portions of each channel were discarded.
In most channels there exists a narrow dark stripe of decreased flux
in the spectra running in the dispersion direction. This stripe has been
known as the ``worm'' and it can attenuate as much as
50\% of the incident light in the affected portions of the
spectrum; - this is due to shadows thrown by the wires on the grid
above the detector.  
Because of the temporal changes in the strength and position of the 
``worm'', CALFUSE cannot correct target fluxes for its presence. 
Therefore, we carried out a visual inspection of the {\it{FUSE}} channels to
locate the worm and we {\it{manually}} discarded the 
portion of the spectrum affected by the worm.
We combined the individual exposures and channels to create a
time-averaged spectrum weighting
the flux in each output datum by the exposure time and sensitivity of the
input exposure and channel of origin. 

\subsubsection{The {\it{FUSE}} Lines} 

The {\it{FUSE}} spectra of DNs in quiescence exhibit mainly 
absorption lines from the WD itself, as the exposed WD  
is the main FUV component of the system. A second FUV component is
sometimes present as a flat and featureless continuum contributing
in the very short wavelengths of {\it{FUSE}} 
($\lambda < 970$\AA\ ). This second component could be  
a hot region of the accretion disk (the inner disk) or the
boundary/spread layer. 
Sharp absorption lines from circumstellar (or circumbinary)
material and/or the ISM are also often seen and,  
if the system is a weak {\it{FUSE}} source, sharp emission lines from air 
glow (geo- and helio-coronal in origin) are present 

Since the WD is the main component of the FUV spectra of DNs in
quiescence, the main characteristic of the {\it{FUSE}} spectra 
is the broad Ly$\beta$ absorption feature due to the gravity 
($Log(g)\approx 8$)
and temperature ($ 15,000$K$<T<25,000$K) of the exposed WD
(in the disk this feature is usually smoothed out due to velocity
broadening, unless the disk is almost face-on).
The other most common absorption features observed
in the spectra of DNs WDs are due to  
C\,{\sc iii} (1175 \AA ),   
C\,{\sc ii} (1066 \AA ),   
Si\,{\sc iii} ($\approx$1108-1114 \AA\ and  
$\approx$1140-1144 \AA ), and 
N\,{\sc ii} (1085 \AA\ when not contaminated by air glow).   
At higher temperatures (T$>25,000$K), 
as the continuum rises in the shorter wavelengths,
the higher orders of the Lyman series also become visible; however,  
they become narrower. At these temperatures the S\,{\sc iv} (1073 \AA ) 
absorption line starts to appear, and, as    
there is more flux in the shorter wavelengths, 
the  C\,{\sc ii} (1010 \AA ) absorption line also becomes visible.
At still higher temperature (T$>50,000$K), the  
C\,{\sc ii} and  Si\,{\sc iii} lines disappear, and the spectrum becomes
dominated by high order ionization lines such as 
N\,{\sc iv} ($\approx$923 \AA ),  S\,{\sc vi} (933.5 \& 944.5 \AA ),  
S\,{\sc iv} (1063 \& 1073 \AA ),  Si\,{\sc iv} (1066.6 \AA ),  
and O\,{\sc iv} (1067.8 \AA ).  

On top of the spectrum of the WD, broad emission lines are found in
quiescent DN systems, usually the O\,{\sc vi} doublet and C\,{\sc iii}
(977\AA\ and 1175 \AA ). These, together with broad emission lines from
N\,{\sc iv} ($\approx$923 \AA ) and  S\,{\sc vi} (933.5 \& 944.5 \AA )  
are often observed in 
nova-like systems (e.g. such as AE Aqr, V347 Pup, DW UMa, see the MAST
{\it{FUSE}} archives) and are not usually present in low-inclination
DN systems in quiescence. This implies that the gas could possibly be 
heated by a shock (e.g. in the boundary layer) and the broadening 
and variation of the lines  suggests they are originating in the disk.  
This scenario has recently been supported by \citet{kro07} 
who for the first time modeled the broad H and He emission lines 
{\it ab initio} as irradiation of the inner disk by the hot boundary
layer and/or white dwarf. One fully expects that the metallic emission 
lines form in the same way. 
The DN system here that exhibits the strongest emission
lines is EK TrA (i=58$^o$). RU Peg (i=33$^o$)
does have some strong emission lines too, but lacks a STIS spectrum
for complete comparison; however, its {\it IUE} spectrum
clearly show strong emission lines such as C\,{\sc iv} (1550\AA)
and Si\,{\sc iv} (1400\AA).   \\ 

The {\it{FUSE}} spectra of the DN systems (and CVs in general) often 
show some ISM molecular hydrogen absorption, which appears 
as sharp lines at almost equal intervals (12\AA ) 
starting at wavelengths around 1110\AA\ and continuing
towards shorter wavelengths all the way down to the hydrogen cut-off around
915\AA\ . In the affected {\it{FUSE}} spectra, we identified the 
most prominent molecular hydrogen absorption lines by their band
(Werner or Lyman), upper vibrational level (1-16), and rotational transition
(R, P, or Q) with lower rotational state (J=1,2,3).

\subsection{The {\it{HST}}/STIS Archival Spectra} 

\subsubsection{Processing the {\it{HST}}/STIS Archival Data} 

All the STIS spectra were processed with
CALSTIS version 2.19. 
Except for the spectrum of EK TrA (obtained in TIME TAG operation mode), 
all the spectra were obtained in ACCUM operation mode.  
All the STIS spectra used the FUV MAMA detector and all were
centered on the wavelength 1425\AA\ . 
The STIS snapshots (with an exposure time of about 600-700s, for 
EY Cyg and SS Aur) were obtained through
the 52x0.2 aperture using the G140L optical element. 
These snapshots consist of one spectrum.
The STIS spectra of VW Hyi and EK TrA (with an exposure 
time of 4,000s and above) were obtained through the 0.2x0.2 aperture
using the E140M optical elements, and with 42 echelle spectra each, 
are of a much higher resolution than the snapshots. Towards the longer
wavelengths, the echelle spectra do not overlap and five gaps are apparent
around $\lambda \approx $1634\AA , 1653\AA , 1672\AA , 1691\AA , and
1710\AA . 

\subsubsection{The {\it{HST}}/STIS Lines} 

The STIS spectra of quiescent DNs are also dominated by the WD and their 
main characteristic is the broad Ly$\alpha$ absorption feature at 
$\approx$1216 \AA . 
The other very common absorption lines are the carbon lines 
C\,{\sc i} (1266 \AA, 1561 \AA, 1657 \AA), 
C\,{\sc ii} (1335 \AA), and C\,{\sc iii} (1175 \AA); 
and the silicon lines
Si\,{\sc ii} (1260 \AA, 1300 \AA, 1530 \AA ),
Si\,{\sc iii} (1300 \AA ),
and Si\,{\sc iv} (1400 \AA ).   

The {\it{HST}}/STIS spectrum also exhibits some broad emission lines. The most
prominent ones are C\,{\sc iii} (1175\AA), C\,{\sc iv} (1550\AA), 
Si\,{\sc iii} (1206\AA), N\,{\sc v} (1240\AA), He\,{\sc ii} 
(1640\AA\ Balmer $\alpha$), and Si\,{\sc iv} (1400\AA).
With their broad emission lines, both the {\it{FUSE}} 
and {\it{HST}}/STIS spectra show evidence of hot gas. 

\subsection{Preparation of the Spectra} 

For four objects (EY Cyg, SS Aur, VW Hyi and EK TrA) we wish to
combine the {\it{FUSE}} spectrum with the {\it{HST}}/STIS spectrum in order to
have a larger wavelength coverage and improve the
synthetic spectral modeling. 
 
However, for each object, before combining the {\it{FUSE}} spectrum with the 
STIS spectrum we must make sure that 
(i) the system was observed in the same 
(high or low) state; (ii) the shape of the spectra (continuum
and lines) is similar (if the spectra are not too noisy); 
(iii) the flux level in the {\it{FUSE}} and 
{\it{HST}}/STIS spectra do not differ 
more than about 50\% ({\it{FUSE}} and 
{\it{HST}}/STIS are very different instruments,
we do not expect their flux level to match exactly); 
(iv) 
for long period systems for which only short exposure time spectra exist, 
the {\it{FUSE}} and STIS spectra have most probably been obtained at a 
different binary orbital phase and it is not clear one can combine
such spectra.  We address these points here in detail as follows. 

(i) 
The observations listed here were not coordinated and consequently some
were carried out during early quiescence while others were carried out 
in late quiescence. For EY Cyg, the outburst recurrence time is 
believed to be about 2000 days and the 
{\it{FUSE}} and STIS spectra were obtained
in deep quiescence. 
EK TrA has also a long outburst recurrence time and it was
observed in deep quiescence (45 days and 155 days after outburst).  
For SS Aur, both spectra were obtained about a month
into quiescence. For VW Hyi the {\it{FUSE}} spectrum was obtained 11 days into
quiescence while the main STIS spectrum was obtained 15 days into
quiescence (we discuss VW Hyi in some more detail in the results 
section). 

(ii) 
For all the systems, a visual analysis shows that the shape of 
the continuum and lines is similar in the {\it{FUSE}} and STIS spectra.  

(iii) 
0nly 2 systems (EY Cyg and SS Aur) have
{\it{FUSE}} and STIS flux levels that do not match exactly. 
For EY Cyg the STIS spectrum
had to be multiplied by 1.5 to match the flux level of the continuum
in the {\it{FUSE}} spectrum, while for SS Aur it was multiplied by 0.93. 
At temperatures of about 25,000K a change of 2,000K can produce a
change of 50 percent in the flux; the accuracy of the temperature
we obtained for EY Cyg is therefore of this order (i.e. $\pm$2,000K). 

(iv) 
Since the systems were observed in quiescence, we expect mainly to 
see the exposed WD with basically no contribution or occultation 
from the accretion disk  (none of these systems is eclipsing).  
The spectra, whether observed at a given phase or averaged
over an entire orbit, should not differ significantly. Only for spectra
with a high S/N ratio and a high inclination do we expect the absorption
and emission lines to exhibit some red- or blue- shift if they were obtained 
at a given orbital phase, while the orbit-averaged spectra will have broader
absorption and emission features. 

Except for RU Peg, all the {\it{FUSE}} spectra exposure times cover a 
significant fraction of the orbital period (and in some cases several 
periods); therefore, the {\it{FUSE}} spectra are averaged over a significantly 
large fraction of the orbit. The STIS spectra of VW Hyi and EK TrA 
are also averaged over the orbit, and for these two systems one can
therefore combine the {\it{FUSE}} and STIS spectra.

The STIS snapshots of EY Cyg and SS Aur are very short  
(less than 1ksec) and  were taken at a particular  
orbital phase. One might therefore argue that the STIS and {\it{FUSE}}
spectra of EY Cyg and SS Aur cannot be combined because of that. 
However, for both EY Cyg and SS Aur,  
they fit except that the absorption lines in the STIS
range are not as deep as in the {\it{FUSE}} range.
The effect of observing the systems at a particular
orbital phase is not very pronounced, and 
this might be due to the lower resolution of 
the STIS snapshots (in comparison to VW Hyi and EK TrA's STIS spectra) 
and the moderate inclination of EY Cyg and SS Aur.
Therefore, for EY Cyg and SS Aur, 
we also combine the {\it{FUSE}} and STIS spectra without concern about
the orbital phase.  

The only object for which we find 
a significant shift in the lines is the long period system
RU Peg, for which we only have a {\it{FUSE}} spectrum of 2,800s. 

Finally, we deredden the spectra according to the E(B-V) values given in 
\citet{ver87,lad91,bru94}. 

\section{Spectral Modeling} 
 
\subsection{The Synthetic Stellar Spectral Codes}

We create model spectra for
high-gravity stellar atmospheres using codes 
TLUSTY and SYNSPEC\footnote{
http://nova.astro.umd.edu; TLUSTY version 200, SYNSPEC version 48} 
\citep{hub88,hub95}. 
Atmospheric structure was computed (using TLUSTY) assuming a H-He LTE 
atmosphere; the other species were added in the spectrum synthesis
stage using SYNSPEC. 
For hot models (say $T>50,000$K) we switched the approximate NLTE treatment 
option in SYNSPEC (this allows to consider and approximate NLTE treatment
even for LTE models generated by TLUSTY). 
We generate photospheric models with effective
temperatures ranging from 12,000K to 75,000K in increments of 
about 10 percent (e.g. 1,000K for T$\approx$15,000K and 5,000K for 
T$\approx$70,000K).
We chose values of $Log(g)$ ranging between 7.5 and 9.5
for consistency with the observed mass. We also varied the
stellar rotational velocity $V_{rot} sin(i)$ from $100$km$~$s$^{-1}$
to $1000$km$~$s$^{-1}$ in steps of $100$km$~$s$^{-1}$ (or smaller if
needed). In order to
try and fit the absorption features of the spectrum, we also vary
the chemical abundances of C, N, S and Si.
For any WD 
mass there is a corresponding radius, or equivalently one single value of
$Log(g)$ (e.g. see the mass radius relation
from \citet{ham61} or see \citet{woo90,pan00} for different
composition and non-zero temperature WDs).

Our suite of stellar spectra generator codes has been implemented and 
includes also the treatment of the quasi-molecular satellite Lyman lines.  
The satellite hydrogen lines appear as strong absorption features 
near 1400 and 1600\AA\  (Ly$\alpha$, in the IUE and STIS range;
\citet{koe85,nel85}) and are somewhat weaker near 1060 and 1078\AA\ 
(Ly$\beta$, in the {\it{FUSE}} range; e.g. \citet{dup06}). 
Following \citet{dup03,dup06}, we used opacity tables computed by 
\citet{all98,all94,all04a} to take into account the quasi-molecular satellite
line opacities of H$_2$ and H$_2^+$ for 
Ly$\alpha$, Ly$\beta$, and Ly$\gamma$.    
The quasi-molecular lines are expected to appear when the temperature
of the WD is $\approx$20,000K or lower, though the gravity, pressure and
magnetic field of the WD can also play a role in the formation of
these lines (larger electronic density will favor recombination,
see e.g. \citet{dup03,gan06} for a more complete discussion).  

\subsection{Modeling the ISM Hydrogen Absorption Lines} 

For all the systems showing ISM atomic and molecular hydrogen 
absorption lines, we identify these lines in the figures to
avoid confusing them with the WD lines. For RU Peg, however,
some ISM lines are deep and broad and we decided 
to model them, especially since some of the WD lines (such as 
S\,{\sc iv} $\lambda  \lambda$1062.6 \& 1073)  are located 
at almost the same wavelengths.   

For RU Peg only, we model the ISM hydrogen absorption lines to assess
the atomic and molecular column densities. This enables us to 
differentiate between the WD lines and the ISM lines, and
helps improve the WD spectral fit. 
The ISM spectra models are generated using a program developed by 
P.E. Barrett.     
This program uses a custom spectral fitting package to estimate the temperature
and density of the interstellar absorption lines of atomic and
molecular hydrogen.  The ISM model assumes that the temperature, bulk
velocity, and turbulent velocity of the medium are the same for all
atomic and molecular species, whereas the densities of atomic and
molecular hydrogen, and the ratios of deuterium to hydrogen and metals
(including helium) to hydrogen can be adjusted independently. The
model uses atomic data of \citet{mor00,mor03} and molecular data of
\citet{abg00}. The optical depth calculations of molecular
hydrogen have been checked against those of \citet{mcc03}.

The ratios of metals to hydrogen and deuterium to
hydrogen are fixed at 0 and $2 \times 10^{-5}$, respectively,
because of the low
signal-to-noise ratio data.  The wings of the atomic lines are used to
estimate the density of atomic hydrogen and the depth of the
unsaturated molecular lines for molecular hydrogen.  The temperature
and turbulent velocity of the medium are primarily determined from the
lines of molecular hydrogen when the ISM temperatures are $< 250$K.

The ISM absorption features are best  modeled and displayed when the
theoretical ISM model (transmission values) is combined with a
synthetic spectrum for the object (namely a WD synthetic spectrum).

\subsection{Synthetic Spectral Model Fitting}

Before carrying out a synthetic spectral fit of the spectra,
we masked portions of the spectra with strong emission lines,
strong ISM molecular absorption lines, detector noise and air glow.
These regions of the spectra are somewhat different for each object and 
are not included in the fitting. The regions excluded from the fit
are in blue in Figures 1, 3, 5, 7, and 10. 
The excluded ISM quasi-molecular absorption lines are 
marked with vertical labels in figures 2, 4, and 9.   
For RU Peg, we model the ISM absorption features and the WD separately; 
namely, we also mask the ISM absorption features when we  model the WD. 

After having generated grids of models for each target, 
we use FIT \citep{numrec}, a $\chi^2$ minimization routine,
to compute the reduced $\chi^{2}_{\nu}$ 
($\chi^2$ per number of degrees of freedom) 
and scale factor values for each model fit.  
While we use a $\chi^2$ minimization technique, we do not 
blindly select the least $\chi^2$ models, but we examine the models 
that best fit some of the features such as absorption
lines (see the fit to the {\it{FUSE}} spectrum alone) 
and, when possible, the slope of the wings of the broad Lyman
absorption features. We also select the models that are in agreement
with the known distance of the system.  

However, in the model fitting, for a given
WD mass and radius, the resulting temperature depends mainly on the 
shape of the spectrum in the {\it{FUSE}} range.
The flux level at 1000\AA\ (between Ly$\delta$ and Ly$\gamma$) 
is close to zero for temperatures below 18,000K, at 30,000K it is about
50\% of the continuum level at 1100\AA\ and it reaches 100\% for T$>$45,000K.  
At higher temperature (T$>$50,000K) the spectrum becomes pretty flat 
and there is not much difference in the shape of the spectrum 
between (say) a 50,000K and a 80,000K model.  
When fitting the shape of the spectrum in such a manner, 
an accuracy of about 500-1,000K is obtained, due to the S/N.
In theory, a fine tuning of the temperature (say to an accuracy of about
$\pm$50K) can be carried
out by fitting the flux levels such that the distance to the system
(if known) is obtained accurately. However, the fitting to the
distance depends strongly on the radius (and therefore the mass)
of the WD. In all the systems presented here the error on the measured mass
of the WD is so large (see Table 1) that it produces 
an error in the temperature much larger than 1,000K. This is
because the Ly$\alpha$ and Ly$\beta$ profiles  depend on both
temperature and gravity. For example \citet{gan01} derived a best fit 
temperature for EK Tra  $T_{eff}\approx 2360\times Log(g)-95$,
where $g$ is the (unknown) surface gravity of the WD 
($7.0 < Log(g) < 9.0$). Additional errors are further
introduced as the distance and reddening are 
rarely known accurately, therefore increasing significantly
the inaccuracy of assessing the temperature by scaling the synthetic flux
to the observed flux. 
For these, we decided to fit the temperature of each model, 
based on the shape of the {\it{FUSE}} spectrum in the shorter wavelengths, for 
a temperature accuracy of only about 5\%, namely accurate
to within 500K for T$< 20,000$K; 1,000K for 20,000K$<$T$<$35,000K;  
and 2,000K for T$>$40,000K (see the Results section). While the
error in temperature is only 500K for a 18,000K model and 2,000K for a 50,000K
model, the relative error $\Delta T/T$ is the same for all the models. 

For all models we first tried fitting solar abundance models and
then tried changing the abundances of some species to improve the
spectral fit. 
In particular, for each system we fit obvious absorption features 
which can determine the abundance of some specific element
(C, Si, S, N).
For example, the C\,{\sc ii} ($\approx$1065 \AA\ ), the Si\,{\sc iii} 
($\approx$1110 \AA\ ) and the Si-C blend ($\approx$1138-1146 \AA\ ) 
are the main absorption features in the {\it{FUSE}} range of a T=20,000K
WD that help assess the Si and the C abundances (these are discussed
in detail for each system in the next section).  

The WD rotation ($V_{rot} sin(i)$) rate is determined by fitting the
WD model to the spectrum while paying careful attention to the 
line profiles in the {\it{FUSE}} portion of the combined spectrum.
We did not carry out separate fits to individual lines but
rather tried to fit the lines and continuum in the same fit while
paying careful attention to the absorption lines. 

It is important to note that, in the modeling, 
the depth of the absorption 
features depends not only on the abundances
but also on the rotational velocity. Increasing the rotational
velocity decreases the depth of the 
absorption features, but broadens the wings - the net effect
being that higher metal abundances result with faster rotational
velocity. With sufficiently high S/N data,  
it becomes possible to assess the rotational velocity and abundances
independently. 

\section{Results and Discussion} 

All the results are presented in Table 5. For some systems 
for which the mass of the WD is relatively unknown, 
we list more than one optimized fit result,  assuming different values
of $Log(g)$ (i.e. for each value of $Log(g)$ we find an optimized fit). 
In the first column we list the name of the system;
in the second column the assumed surface gravity on a logarithmic  
scale $Log(g)$  
(where it is understood that $Log$ is the logarithm to the base 10 
- as opposed to $log$ which denotes the logarithm to the base $e$) 
in cgs units;  
in column (3) we list the effective surface temperature we 
obtained for the WD; in column (4) we list the projected rotational
velocity which was obtained when matching the stellar absorption 
lines; in columns (5), (6), (7) and (8) we list the abundances
of Carbon, Silicon, Nitrogen and Sulfur (respectively) in solar
units; in column (9) we list the distance we obtained from the
fitting; in column (10) we list the parameter by which the STIS
flux had to be multiplied to match the {\it{FUSE}} flux; in column (11)
we give the $\chi^2_{\nu}$ value; and in column (12) we indicate 
the number of the figure displaying the model fit. 

\subsection{EY Cygni} 

EY Cygni is a  peculiar long period U Geminorum-type of DN,         
with a very massive accreting WD ($M_1 = 1.26M_{\odot}$ or possibly
larger), an outburst amplitude of 4.1 magnitude with a recurrence
time of 2000 days, and an orbital period of 11.0238 hrs  
\citep{smi97,cos98,tov02,cos04}.  
H$\alpha$ and S\,{\sc ii} ground-based imaging has revealed it is
associated with a non-homogeneous shell with a size of about 25arcsec
\citep{tov02,sio04b}.  
It is possible that this shell is the result of a recent nova explosion.
The spectral type of the secondary is not certain, it is believed to 
be a K5V to M0V star, from which the lower limit of the distance to 
EY Cyg has been inferred to be at least 250pc. Since EY Cyg is
not embedded in the background Cygnus superbubble \citep{boc85}, 
the upper limit for its distance is 700pc. An additional characteristic
which makes EY Cyg a peculiar DN is its anomalous 
N\,{\sc v}/C\,{\sc iv} 
ratio. \citet{win01} first reported a very strong 
N\,{\sc v} emission and a very weak (or absent) 
C\,{\sc iv}, which is atypical of most DNs in which
C\,{\sc iv} is usually the most prominent and common emission line in
their FUV spectra. More recently \citet{gan03} reported an analysis 
of {\it{HST}}/STIS snapshots of 4 CVs (including EY Cyg) with
anomalously large 
N\,{\sc v}/C\,{\sc iv} 
line flux ratios, similar to those 
observed in AE Aqr. So far 10 systems have been identified with
such an anomalous 
N\,{\sc v}/C\,{\sc iv} 
ratio. 

EY Cyg is an extremely weak {\it{FUSE}} source and the spectrum is therefore
very noisy, which is the reason we binned its {\it{FUSE}} spectrum 
at 0.5\AA\  for the fitting (Figure 1). The STIS snapshot spectrum was binned at
the default value of 0.58\AA .  
Consequently the relative weight of the {\it{STIS}} spectrum is about
twice as large as that of the {\it{FUSE}} spectrum. 
In preparation for the model fitting 
we had to multiply the STIS flux by 1.5 to match 
the {\it{FUSE}} flux level; this change of 50\%
in flux corresponds to a $(0.5)^{\frac{1}{4}}\approx$ 10\% in the 
value of $T$. Therefore, the relative error in temperature in our spectral 
modeling of the combined {\it{FUSE}}+STIS spectrum of EY Cyg 
is of the order of $\sim 0.1$ {\it ab initio}.  
Since the distance to the system is relatively unknown, 
there is no restriction on the flux level to minimize the
number of best fits to one in the $Log(g)-T_{eff}$ parameter
space. Therefore, we restricted  
the modeling assuming only two different values for the mass of
the WD.  Because of the extremely large WD mass of the system 
(which within the
margin of error well exceeds the Chandrasekhar mass limit for 
a WD)  we ran models assuming $Log(g)=9.0$ (corresponding to 
$M_{wd} = 1.21M_{\odot}$) and $Log(g)=9.5$ (corresponding to 
$M_{wd}=1.35M_{\odot}$).  
 
Within the range of the temperature error (10\%) the best fit models
spread between about T=27,000K and T=33,000K. 
However, the high temperature
($T>32,000$K) models do not produce enough flux in the longer wavelengths
of STIS, while the lower temperature ($T<28,000$K) models do not have 
enough flux in the shorter wavelengths of the {\it{FUSE}} 
portion of the spectrum. 
Because of that we rejected the $T=27,000$K and $T=33,000$K models.
In addition, in the observed STIS spectrum the bottom of the  Ly$\alpha$ does not
go to zero and this corresponds to models with $T\sim 30,000$K and higher. 
Because of the difference in fluxes between STIS and {\it{FUSE}} we decided to
check how the results would change if we fit the spectra separately. 
We found that the STIS spectrum gave the same temperature, however with
a distance about 20 percent larger as the main driving elements in choosing 
the best fit model were the same, namely: the Ly$\alpha$ profile and
the longer wavelengths of STIS. 
Fitting the {\it{FUSE}} spectrum alone gave a
temperature about 2,000K higher than for the combined spectrum. 
We therefore inferred for EY Cyg $T=30,000$K$ \pm 2,000$K assuming 
$Log(g)=9.0$. The 2,000K error in $T$ produces an error of $\pm 60$pc 
in the distance as listed in Table 5. This 30,000K model is shown in Figure 1. 
In Figure 2 the 30,000K WD model assuming $Log(g)=9.0$ 
is shown with the {\it{FUSE}} spectrum  binned here at 
0.1 \AA\ to show the sharp emission and absorption lines.  

We also checked how the inclusion of the quasi-molecular hydrogen 
affects the result and found that it does not improve the fits.  

Next, we assume $Log(g)=9.5$ to check the effect of larger WD mass.  
Using the same considerations as before we found that the best fit models
range between $30,000$K and $34,000$K, or about 2,000K higher than the
$Log(g)=9.0$ models. 

The projected rotational
velocity we obtained for all the models is relatively low: 100km$~$s$^{-1}$.  
We confirm the anomalously low Carbon abundance, namely we found
that the best Carbon line fits are obtained for abundances ranging between
0.01 and 0.05 solar. 
The low carbon abundance was set to match the 
{\it{FUSE}} absorption features of 
C\,{\sc iii} 1175 \AA ,  
C\,{\sc ii} 1066 \AA , and 
C\,{\sc i}-C\,{\sc iii} at 1128 and 1140 \AA . 
There is also a complete absence
of C\,{\sc iii} emission at 977\AA\ and 1175\AA , 
and C\,{\sc iv} emission at 1550\AA\ .   
There is a strong and broad   N\,{\sc v} 1240 \AA\ emission line.
The Nitrogen abundance ($2 \times$ solar) was obtained by fitting the
N\,{\sc i} (1135\AA),     
N\,{\sc ii} (1185\AA), and  
N\,{\sc iii} (1189\AA) lines. However, these lines are not very pronounced
and are affected by air glow emission.   
The Sulfur abundance ($6 \times$ solar) was obtained by fitting the
S\,{\sc iv} (1073\AA) and 
S\,{\sc i} (1097\AA ) lines.  
The Silicon abundance ($0.6 \times$ solar) was obtained by fitting the 
Si\,{\sc iv} (1125\AA), and 
Si\,{\sc iii} (1140-1145\AA) lines.  \\  

\citet{sio04b} modeled the combined {\it{HST}}/STIS and {\it{FUSE}}
spectra of EY Cygni. Despite the noisy, weak {\it{FUSE}} 
spectrum, they managed to be obtain two possible solutions which in a 
statistical ($Chi^2_{\nu}$) sense differ little from each 
other. If the combined FUV spectra was due to a single temperature white 
dwarf alone, then for their adopted mass of $1.2 M_{\odot}$ 
(assuming $Log(g)= 9$) with no interstellar reddening (see below), then the WD 
has $T_{eff} = 24,000$K. However, the combined {\it{FUSE}} plus 
STIS spectra also yielded a similar statistical fit if a two-temperature 
white dwarf model was applied to the data. 
This fit corresponded to a cooler, slowly rotating WD photosphere 
($T_{eff} = 22,000$K) with the second temperature 
($T_{eff} = 36,000$K) representing a putative hot, rapidly rotating, 
equatorial accretion zone (an accretion belt). 
We note here that their noisy {\it{FUSE}} spectrum showed a sharp increase in 
flux extending even beyond the Lyman limit. 
This extra flux is an artifact of the CaLFUSE background 
subtraction and, therefore, in the present woork we process the {\it{FUSE}} 
data with latest (and final) version of 
CalFUSE (v3.2) which lowered the flux at wavelengths < 950\AA , 
removing the shortward flux excess. 
The new combined {\it{HST}}/STIS plus reprocessed {\it{FUSE}}
data is fit in the 
present paper with the {\it{FUSE}} spectral region shortward 
of 950\AA masked in the fitting. 

Our single WD component best fit, assuming $Log(g)=9.0$, 
has $T_{eff} = 30,000$K, with a distance $d\sim$500pc; and assuming
$Log(g)=9.5$ we found a WD temperature of $32,000$K
with a distance of $\sim$300pc. 
EY Cyg does not show any sign of quasi-molecular satellite lines, 
and this is consistent with our higher temperature model 
(see the Discussion section),   
and no need to add a second component 
since this component was driven by the flux excess which the 
re-processing removed. 

\subsection{SS Aurigae} 

SS Aur is a U Gem type DN with 
an orbital period $P_{orb}$=4.3872 hrs \citep{sha86},  
a WD mass $M_{wd} = 1.08\pm0.40$ $M_{\odot}$, 
a secondary mass $M_{2} = 0.39\pm 0.02$ $M_{\odot}$, and
a system inclination $i = 38^o\pm16^o$ \citep{sha83}. SS Aur
has an {\it{HST}} FGS parallax measurement of 497 mas \citep{har99} giving
a distance of 201pc.

\citet{lak01} analyzed the IUE archival spectra of 
SS Aur; their best-fit model photosphere has T$_{eff} =
30,000$K, $Log(g)=8.0$, and solar composition abundances; while the
best-fit
accretion disk model has M$_{wd} = 1.0$ $M_{\odot}$, $i = 41^{o}$, and
$\dot{M} = 10^{-10}$ $M_{\odot}yr^{-1}$. 

Using a newly available Hubble FGS parallax, 
\citet{lak01} showed for the first time that the FUV flux 
of SS Aur during quiescence was dominated by, if not provided entirely 
by, a hot WD, not an accretion disk. This was 
something of a milestone because the prevaling view was 
that the underlying degenerate was detected in the FUV only in those
dwarf novae which clearly revealed the broad 
Ly$\alpha$ profile of a WD, namely U Gem, VW Hyi and
WZ Sge. 

More recently \citet{sio04a} analyzed the {\it{FUSE}} spectrum of the system. 
Their best fit model is a 27,000K WD contributing 73\%
of the flux with a hot belt (48,000K) contributing the remaining 27\%
of the flux, with a corresponding distance of 267pc. 
However, a single WD component also gives a good fit
with a temperature $T_{wd}$=33,000K and a distance of 303pc.  
They also suggested that the absence of the
C\,{\sc iii} (1175\AA) absorption line might be 
a sign of composition deficit  
in carbon. It is important to note here that in these previous
studies the spectra of SS Aur were not dereddened.  

The STIS spectrum of SS Aur was obtained as part of a snapshot
program \citep{gan03}  and the single STIS spectrum is being 
analyzed also in \citet{sio08}.  
Here, we analyzed the combined {\it{FUSE}}+STIS spectrum of SS Aur
which we dereddened assuming E(B-V)=0.08 \citep{bru94}. 
The {\it{FUSE}} spectrum of SS Aur was obtained 28 days after outburst, 
while the STIS spectrum was obtained 34 days after outburst. 
The fit was carried out with a binning of 0.2 \AA\ in the {\it{FUSE}}
spectrum and 0.58 \AA\ in the {\it{STIS}} spectrum, consequently
the relative weight of the {\it{FUSE}} spectrum is about 50\% larger
than that of the STIS spectrum. 

The best fit model for a $\approx$1.1$M_{\odot}$ ($Log(g)=8.71$) mass WD
and a distance of 200pc gives a temperature of 31,000K, solar abundances, 
a rotational velocity of 400km/s and $\chi^2_{\nu}$=1.476. This
fit is the best when fitting the flux level while keeping the distance  
and radius of the star fixed.
If we relax the restriction on the distance,  
the best fit model with $Log(g)=8.71$ gives a temperature of 
33,000K, a distance of about 240pc and a slightly lower $\chi^2_{\nu}$ (1.472).
The difference in the values of the $\chi^2_{\nu}$ between the 31,000K 
and the 33,000K models is not significant.   
However, it  is this discrepancy between the apparent temperature
of the WD and its flux that has led to the adoption of second model
components in other CVs. However, the WD mass of SS Aur is given
with a large error ($\pm 0.4M_{\odot}$). We, therefore, computed more
models with $Log(g)=8.31$ and $8.93$. 
The $Log(g)=8.31$ model, for a distance of 200pc gave a temperature of 27,000K,
while the lowest $\chi^2_{\nu}$ best fit model for that value of $Log(g)$ gave a temperature
of 30,000K and a distance of 254pc.  
For the $Log(g)=8.93$ model, the distance of 200pc is achieved with a
temperature of 34,000K. This model, however, is also the lowest $\chi^2_{\nu}$ best fit 
model for that value of $Log(g)$ with the least $\chi^2_{\nu}$ 
of all (1.471). 

A model fit to the STIS spectrum alone of SS Aur (which was carried out
in \citep{sio08}) 
gave the same results, namely a 34,000K WD assuming $Log(g)=8.8$.
The only difference with our modeling is that the STIS spectrum 
in \citet{sio08} was NOT normalized to the {\it{FUSE}} spectrum, which
explain the slight difference in the value of $Log(g)$. We then 
ran a model to fit the {\it{FUSE}} spectrum alone and found total 
agreement with the {\it{FUSE}}+STIS spectral fits, namely that the best
fit model is again a T=34,000K WD, spinning at 400km/sec with
solar abundances, for $Log(g)=8.93$ and a distance of 200pc. 
We therefore adopted this model as the best model fit 
for SS Aur, as it is produced within the margin of error of the WD
mass and it is the simplest (one component) model possible. 
We do not see any need to add a second
FUV component in the spectral modeling of SS Aur.
The presence of a second (flat and featureless) component to the
spectrum of SS Aur has not been detected as in the case of VW Hyi, and
in the present work we restrict our modeling to a single WD component.
The $Log(g)=8.93$ best fit model with a temperature of 34,000K 
is shown in Figure 3. In Figure 4, we show the same model superposed to 
the {\it{FUSE}} spectrum binned at 0.1 \AA . 

Usually C\,{\sc ii} (1066\AA) and C\,{\sc iii} (1175\AA) are 
in absorption in a 30,000K WD stellar atmosphere while  
C\,{\sc iii} (1175\AA) and C\,{\sc iii} (977\AA)
can show emission originating in a hot second component (e.g. thin
disk, accretion column, BL). From Figure 4, we see 
C\,{\sc ii} (1066\AA\ and 1010\AA ) is in absorption while 
C\,{\sc iii} (977\AA) is in emission. 
C\,{\sc iii} (1175\AA) on the other hand does not show any sign of absorption 
nor emission (it could be a superposition of absorption and emission). 
We therefore decided to fit the carbon abundances by fitting the 
C\,{\sc ii} lines;  
achieved by simply assuming solar abundances 
(we note that if we assume a low carbon abundance to fit the C\,{\sc iii}
1175 feature, then the C\,{\sc ii} absorption features at 1010\AA\ and
1066\AA\ are not fitted) . We do not find
evidence that the WD is deficient in carbon as is the case for
EY Cyg, especially since the C\,{\sc iv} (1550\AA) is
in strong emission in the STIS snapshot. 

In the present work, we have improved the line 
identifications and the fitting in the shorter wavelengths
without the need for a second component. The second component 
\citep{sio04a} gives too much flux in the short wavelengths.
The distance we obtained is in very good agreement with the
distance estimated from the parallax, which is smaller by 50\% than the
distance obtained in \citet{sio04a} (a result of our dereddening). 

\subsection{VW Hydri} 

VW Hyi is a key system for understanding DNs in general.
It is one of the closest \citep[placed it at 65 pc]{war87}
and brightest examples of an SU UMa-type DN. It lies
along a line of sight with a low interstellar column
\citep[estimated the H\,{\sc i} column density to be
$\approx 6 \times 10^{17}$ cm$^{-2}$]{pol90},
which has permitted study of VW Hyi in nearly all wavelength ranges,
including detection in the usually opaque extreme ultraviolet
\citep{mau96}. For these reasons, it is one of the best-studied 
systems. Coherent and quasi-coherent soft X-ray oscillations 
have been detected, and its X-ray luminosity has been found to be 
unexpectedly low \citep{bel91,mau91}.
Its boundary layer (BL) has been shown to emit most of its
energy in the FUV band \citet{pan03,pan05,god05}.   
VW Hyi is near the lower edge of the CV period gap,   
with an orbital period of 1.78 hrs and a quiescent
optical magnitude of 13.8.
It is a member of the SU UMa class of DNs, which undergo
both normal DN outbursts and superoutbursts. The normal
outbursts last 1-3 days and occur every 20-30 days,
with  peak visual magnitude of 9.5. The superoutbursts
last 10-15 days And occur every 5-6 months, with peak
visual magnitude reaching 8.5. The mass of the accreting
WD was estimated to be 0.63 $M_{\odot}$ \citep{sch81},
but more recently a gravitational redshift determination
yielded a larger mass $M_{wd}=0.86 M_{\odot}$ \citep{sio97}.
The inclination of the system is $\approx 60^o$ \citep{hua96a,hua96b}.

VW Hyi was first observed at UV wavelengths with {\it IUE}.  Based on
observations of VW Hyi in quiescence, which showed a broad absorption
profile centered on Lyman $\alpha$, \citet{mat84} argued that
UV light from the system was dominated by the WD with
$T_{eff} = 18,000 \pm 2,000$ K
(for $Log(g)=8$). Much higher S/N spectra were obtained
with {\it HST}.  In particular, \citet{sio95a}
found the WD temperature to be 22,000K, 
the first metal abundances, with evidence of CNO processing of material 
in the WD photosphere. They also 
derived the first rotational velocity for a WD in a CV using the GHRS 
G160M spectrum of VW Hyi \citep{sio95b} 
thus showing that weak boundary layer emission from some CVs could not 
be explained by their WDs rotating near 
Keplerian speeds. \citet{sio96,sio01} concluded that the 
temperature of the WD varied by at least 2000K depending on the time 
since the outburst. They also presented evidence 
for rapidly rotating "accretion belt" in the system. The
accretion belt was to be understood physically as a region of the WD
surface spun up by accreting material with Keplerian velocities.
Furthermore, \citet{sio97}, using the GHRS on {\it HST}, measured
the gravitational redshift of the WD and concluded that the mass of the
WD was $0.86 ^{+0.18}_{-0.32} M_{\odot}$ and that the rotation rate of
the WD was $\sim$ 400 km~s$^{-1}$.  All of these observations were limited 
to a wavelength range longward of about 1150 \AA. However, an 820-1840 \AA\
spectrum of VW Hyi was obtained using the Hopkins Ultraviolet Telescope
(HUT).  \citet{lon96} found that the HUT spectrum was reasonably
consistent with a WD with a temperature of about 17,000K, but that an
improved fit to the data at that time could be obtained with a
combination of emission from a WD and an accretion disk.

Higher S/N spectra of VW Hyi in quiescence were obtained more 
recently with {\it{FUSE}}. With a practical wavelength range
of 904-1188 \AA, {\it{FUSE}} is sensitive to a second component, since the
expected flux from a WD with a temperature of about 20,000 K is very
different at 950 and 1100 \AA.  In addition, the {\it{FUSE}} spectral
range and high spectral resolution allows the study of a broad range of
line transitions.
In the present work we only use one {\it{FUSE}} data set 
(B0700201), which presents
only a small contribution from the second component. Since the
second component is being studied elsewhere (Long 2007, private communication) 
using different {\it{FUSE}} data sets - E1140109, E1140112, E1140113), 
and it represents
only a few percent of the flux, we neglect its contribution in the
present work. 
In a previous work \citep{god04} we modeled the {\it{FUSE}} spectrum
of VW Hyi, including a fast rotating hot accretion belt, and obtained
a WD temperature of 23,000K. However in that work we did not deredden
the spectrum, nor did we include the quasi-molecular satellite line
in the modeling, and the {\it{FUSE}} spectrum was not combined with the STIS 
spectrum. 

In the present analysis, 
we included the quasi-molecular lines in the synthetic
spectral modeling, to fit the {\it{FUSE}} spectra better in the wavelength
range $\approx 1060-1080$\AA , and we dereddened the spectrum assuming
E(B-V)=0.01 \citep{bru94}. We are aware that because of the very low
hydrogen column density one actually expects VW Hyi to have a
zero reddening. However, (i) such a small reddening value does not affect
the results significantly, and (ii) we want to be consistent with the
reddening values given in the literature and adopted in this work.
The {\it{FUSE}} spectrum we used here
was obtained 11 days after outburst, while 
there were three STIS spectra in the archive which we combined 
(weighting them by exposure time) to improve the S/N. 
Two short exposure STIS spectra were obtained 1 and 7 days
after outburst; however, they are actually very similar to the long 
exposure STIS spectrum obtained 15 days after outburst. 
The {\it{FUSE}} and STIS spectra do match and fit well together.

Since the WD mass might be somewhere between $\approx 0.6$ and $0.9 M_{\odot}$,
we computed models for two different values of $Log(g)$: $8.0$ and $8.5$. 
The model with $Log(g)=8.0$ was in better agreement with the distance
of the system ($\sim 60$pc)
and gave a temperature of 22,000K, a projected rotational
velocity of 400km/s and abundances as specified in Table 5.
This model is shown in Figures 5 \& 6. In both figures the {\it{FUSE}} and
STIS spectra are binned at 0.1 \AA , and the relative weight of the
STIS spectrum is a little above twice that of the {\it{FUSE}} spectrum.  
Due to the
excellent fit of the Si, C and N lines, we are confident that 
these abundances are accurate. While the model does not fit the 
flux (from a second component) in the short wavelengths of {\it{FUSE}}, it
accurately follows the shape of the continuum and lines at wavelengths
$\lambda \sim 1050$\AA\ all the way into the longer wavelengths of STIS.  

The inclusion of the quasi-molecular opacity did improve the 
fit in the {\it{FUSE}} range 1055\AA - 1183\AA , but did not change the
fit in the STIS spectrum around $\approx 1300-1400$\AA\ range,
because of the temperature of the
model. The WD could be massive with $Log(g) \sim 8.5$ or larger.  

From this fit, it is evident that
a WD alone is not enough to reproduce all the features of the spectrum.
The bottoms of Ly$\alpha$ \& Ly$\beta$ do not go to zero and indicate the
presence of an additional component, which may or may not be 
linked to the broad emission features. 
As  stated earlier, we do not intend in this paper to model the
second FUV component of the system as this is being done elsewhere.
We concentrate here on fitting the main continuum, the absorption 
lines, the quasi-molecular satellite lines and combining the {\it{FUSE}}
spectrum with the STIS spectrum in order to assess the basic parameters
of the WD: temperature, abundances, projected rotational velocity,
and gravity. Our model neglects possible contribution from a
quiescent (and optically thin) accretion disk.  
As an additional test we ran models for the fitting of the STIS
spectrum alone and found complete agreement with the best fit
model for the {\it{FUSE}}+STIS spectrum: a $Log(g)=8.0$ WD with a 
temperature $T=22,000$K and a distance of 60pc, therefore confirming
our result for the combined {\it{FUSE}}+STIS spectrum.  
  
\subsection{EK Trianguli Australis} 

EK TrA is a poorly studied southern DN with an 
orbital period just above 1.3 hrs. 
It belongs to the SU UMa subclass (optical superhump was 
confirmed by \citet{has85}) and     
has often been compared to the well studied DN VW Hyi because of
their similarity. However, EK TrA has a much longer outburst period
than VW Hyi and can therefore easily be observed during deep quiescence
for the study of its WD. Despite this, however, 
the mass of its WD is unknown and its distance is
estimated to be $d>180$pc from the non-detection of its secondary 
\citep{gan97}. It is important to note that while  
\citet{war87} put it at 200pc, \citet{has85} initially 
derived $d\approx 68-86$pc.   

EK TrA was observed in August 1980 with IUE 
and 8 spectra were obtained with the SWP and LWR 
cameras covering mid-outburst and late decline
from the outburst. The WD was revealed and 
contributed about 25\% of the IUE SWP flux, with
an accretion disk contributing the remaining 75\% \citep{gan97}.  
More recently, EK TrA was observed with {\it{HST}} 155 days after
outburst and a  
high resolution STIS spectrum was obtained and analyzed by
\citep{gan01}. They confirmed the temperature of the WD found
in \citet{gan97} (T=18,800K) assuming a canonical mass of 0.6$M_{\odot}$,
derived a rotational velocity of 200km$~$s$^{-1}$ and slightly
sub-solar abundances (0.5). They note that assuming a WD mass in the range 
0.3-1.4$M_{\odot}$ leads to a temperature in the range 17,550-23,400K.  
More recently EK TrA was observed with {\it{FUSE}} 45 days after outburst
and two 33ksec spectra were obtained with the same flux level as the
existing STIS spectrum. The {\it{FUSE}} spectra have not been modeled
previously, and 
we therefore decided to model EK TrA using
the combined {\it{FUSE}}+STIS spectrum. 

Since the WD mass of the system is basically unknown, we ran model
fits for $Log(g)=7.5$, $8.0$, and $8.5$ (corresponding to a mass of about
0.40, 0.65 and 0.93$M_{\odot}$ respectively) and obtained a WD
temperature of 15,000, 17,000 and 18,000K respectively. 
The $Log(g)=8.0$ best fit model is presented in Figure 7. 
The broad emission lines (in blue/dark grey) have been masked
for the fitting and both the {\it{FUSE}} and STIS spectra have
been binned at 0.175\AA\ . In Figure 8 we show the {\it{FUSE}} 
spectrum together with the best fit model shown in Figure 7.    
In the wavelengths $\lambda < 1050$\AA only the broad O\,{\sc vi} doublet 
and the C\,{\sc iii}(977 \AA\ ) emission lines are from the source, 
all the other sharp emission lines are due to air glow. The flux
level at these wavelengths is of the same order as the noise and
we do not attempt to model the spectrum there.  
In order to
fit the line profiles in the {\it{FUSE}} range we had to set C=0.1 solar
and Si=0.6 solar while keeping S and N to their solar values. 
Here also we found that the inclusion of the quasi-molecular
satellite hydrogen line opacity improves the fit significantly,
especially in the {\it{FUSE}} range where the flux suddenly drops shortward
of the nitrogen absorption line at 1184 \AA . In the STIS range this is
not as pronounced but the drop in flux is also apparent shortward
of the Si emission lines (1400\AA ). 
The distance we obtained is at most 137pc ($Log(g)=7.5$), 
somewhere between \citet{has85} estimate and \citet{war87}.  
The temperatures we obtained here are lower than \citet{gan97}.
The differences between our result and \citet{gan97} are probably
due to the fact that, in their analysis, \citet{gan97} did not 
include the opacity of the
quasi-molecular satellite hydrogen lines, and they did not deredden the
spectrum (it has E(B-V)=0.03; \citet{bru94}).
They interpreted the excess of flux in the longer wavelengths
of STIS ($>$1,550\AA ) as possible emission from an optically thin
accretion disk during quiescence, while we found, after
dereddening the spectra, no evidence of flux excess. 
We note here that the complete absence of ISM molecular hydrogen absorption
lines in the {\it{FUSE}} spectrum of EK TrA suggests that the system might indeed
be located rather nearby (say $<$150pc). 
For comparison, in the present work, only the {\it{FUSE}} spectrum of
VW Hyi at 65pc shows such a complete absence,  
while SS Aur at 200pc clearly shows some molecular hydrogen lines.   
We therefore speculate that the system is probably located at about 125pc, 
as supported by our results. 

\subsection{RU Pegasi} 

RU Peg is a U Gem type DN with an orbital period $P_{orb}$ = 8.99 hrs, a system 
inclination $i = 33^{o}$, a secondary spectral type K2-5V, and 
a primary (WD) mass $M_{wd} = 1.29_{-0.20}^{+0.16} M_{\odot}$ 
\citep{sto81,wad82,sha83}. The near-Chandrasekhar
mass for the     WD      has been corroborated by the Sodium (8190\AA)
doublet radial velocity study of \citet{fri90}.  They obtained a mass of
1.24 M$_{\odot}$ for the     WD      and also found very good 
agreement with the range of plausible
inclinations, found in the study by \citet{sto81}. Recently, a Hubble FGS
parallax of $3.55\pm0.26$ mas was measured by \citet{joh03} implying
a distance of 282pc.

\citet{sio02} analyzed 4 IUE spectra of RU Peg obtained in deep
quiescence, and found a WD temperature T$_{eff} = 50-53,000$K 
with a distance of $\sim$ 250 pc, assuming $Log(g)=8.7$ 
($M=1.1M_{\odot}$). More recently, a {\it{FUSE}} spectrum of RU Peg was 
obtained 60 days after outburst. The {\it{FUSE}} observation lasted for 
a little more than 3000s (raw exposure time). 
From all the 5 systems in this analysis, RU Peg is the most affected
by ISM absorption lines. We show the
{\it{FUSE}} spectrum of RU Peg with line identifications in Figure 9, 
where we have marked only the most prominent molecular hydrogen lines. 
Two modelings have been carried out \citep{sio04a,urb06} for the 
{\it{FUSE}} spectrum of RU Peg; and both found
a WD temperature similar to \citet{sio02} assuming $M_{wd}=1.2 M_{\odot}$.

While we are confident that the WD in RU Peg is actually 
massive and that its temperature is very large ($>$50,000K), the {\it{FUSE}}
spectrum has never been modeled in detail as many of the 
ISM molecular hydrogen lines were misidentified.
In addition, since this {\it{FUSE}} analysis was carried out, CALFUSE 
has been updated and we are making use of this implementation.
For RU Peg, we decided to also model the 
ISM molecular and atomic hydrogen absorption lines to help improve
the model fit.

In Figure 10 we show the best-fit model to the {\it{FUSE}} spectrum of RU Peg
(binned at 0.1\AA ).
This model consists of a synthetic WD spectrum model multiplied by 
the transmission values obtained from the ISM model.  
This WD model has a temperature of 70,000K, a rotational velocity of 40km/s,
$Log(g)=8.8$, 
subsolar carbon and silicon abundances and supra solar nitrogen and 
sulfur abundances as follows.  
The silicon abundance was set to 0.2 times solar 
to match the Si\,{\sc iv} (1066 \AA) line. 
The nitrogen abundance was set to solar to match 
the N\,{\sc iii} (980, 1183, \& 1184.5 \AA)  lines.  
The carbon abundance was set to 0.2 times solar to match the feature at 
1108 \AA\ and the flat portion of the spectrum at 1160-1170 \AA .
The sulfur abundance was set to 10 times solar to match the strong 
absorption of the S\,{\sc vi} and S\,{\sc iv}   lines.  
It is worth noting that if the N/C ratio is anomalous in the WD
photosphere then this would imply recent thermonuclear processing 
in the WD, because the disk emission lines do not indicate  
a N/C ratio larger than one, but rather the opposite \citep{sio02}. 
However, our abundances determination is not
robust because the nitrogen and carbon absorption lines are afffected
by broad emission lines and ISM molecular hydrogen absorption lines. 

The projected rotational velocity was found by matching       
the profile of the carbon, oxygen, silicon and sulfur lines,  
and in particular the square-shaped bottom of the 
C\,{\sc iii} 1175\AA\  absorption feature. 
The ISM model (in Figure 10) has zero metallicity, 
a temperature of 100K, 
a turbulent velocity of b=30km$~$s$^{-1}$, 
a molecular hydrogen column density of $1 \times 10^{16}$cm$^{-2}$ and 
an atomic hydrogen column density of $1 \times 10^{17}$cm$^{-2}$.
This ISM model fits best the stronger (molecular hydrogen) absorption lines 
but does fit the sharper and weaker (molecular hydrogen) absorption lines. 

It is worth noting that all the synthetic WD spectra we generate   
between $T\sim 50,000$K up to $T\sim 80,000$K 
(with an approximate NLTE line treatment option switched on in SYNSPEC 
and using LTE stellar atmosphere models from TLUSTY) fit 
the data similarly, the main difference is the 
radius (and therefore mass and gravity) of the WD which is fixed  
by scaling the model to the data assuming a distance of 282pc. 
Consequently, the 50,000K model had $Log(g)=8.55$, 
while the 70,000K model had $Log(g)=8.8$. In order to agree with 
a mass of $1.29M_{\odot}$ we ran models with $Log(g)=9.23$ with  
increasing temperatures and found that the model that agrees with
the distance of 282pc has a temperature of 110,000K. 
At such a high temperature the S\,{\sc iv}, Si\,{\sc iv} 
and O\,{\sc iv} lines between 1060\AA  and 1080\AA are not deep enough 
to match the observed features. In addition,   
since we performed only the simplest NLTE approximation, and
also obtained $\chi^2=0.827$ (against 0.539 for the T=70,000K
model), we decided not to chose the 110,000K  model as a best fit model.  

On the other hand, at 50,000K, the O\,{\sc iv} line and  
the nominally strong N\,{\sc iv} and S\,{\sc vi} lines are not pronounced.  
The temperature had to be increased to 70,000K 
to clearly show the appearance of these lines in the synthetic spectrum
(including the O\,{\sc iv} line at 1067.8\AA ). Therefore, we chose
the 70,000K model as our best fit model.  

However, the Si\,{\sc iv} and 
O\,{\sc iv} lines (between 1066 and 1068 \AA ) are contaminated
by sharp ISM absorption lines from molecular hydrogen  
(L3P2 1066.90\AA, \& L3R3 1067.48\AA) and the 
Ar\,{\sc I} 1066.66\AA\ line  
(the other argon line Ar\,{\sc I} 1048.20\AA\ is also
present). However, these sharp ISM lines were not modeled 
accurately in the ISM model presented in Figure 10.  
Consequently, in order to fit the sharper absorption lines of the ISM we  
generated a second ISM model. This model fits the sharp lines better
but does not fit the stronger lines as well (it is likely that the 
ISM has more than one component). 
The second ISM model has zero metallicity, 
a temperature of 300K, 
a turbulent velocity of b=10km$~$s$^{-1}$, 
a molecular hydrogen column density of $1 \times 10^{16}$cm$^{-2}$ and 
an atomic hydrogen column density of $1 \times 10^{18}$cm$^{-2}$ 
and a blueshift velocity of -35km$~$s$^{-1}$.
In Figure 11 we present the 70,000K WD synthetic spectrum including 
the second ISM model in the region 1060\AA\ - 1080\AA\ . 
From this model we see that the 
O\,{\sc iv} line is distinct from the molecular hydrogen line (L3R3) 
but it is not very strong. 
As for the Si\,{\sc iv} lines, they are contaminated by argon and a molecular
hydrogen line (L3P2). Since argon (and iron, and all the metals) is not
included in the ISM model, we cannot confirm whether the Si\,{\sc iv} line
is present. We also note that S\,{\sc iv} is shifted to the red compared
to the WD model.    
Indeed, a close look at all the lines from the system 
reveals that some of the (broader) lines are 
shifted by as much as 0.5\AA\ to the red; this is especially the case 
for the S\,{\sc vi}, O\,{\sc vi}, S\,{\sc iv} and C\,{\sc iii}
lines. This could be due to the motion of the WD in the binary,
as the red shift is close to the maximum radial velocity shift of
the WD in the system and the {\it{FUSE}} spectrum had a duration much
shorter than the binary period. It is possible that these strong absorption
lines do not form in the stellar photosphere, but possibly in a corona 
above it or in a hot boundary layer in front of it.  
This can be determined only if we have line velocities in 
phase-resolved FUV spectra to compare with the velocity derived 
for photospheric lines in the white dwarf's rest frame.

\subsection{The Quasi-Molecular Satellite Lines} 

One of the improvements we made in the present work is the modeling
of the quasi-molecular satellite lines of hydrogen in the {\it{FUSE}}
spectra of VW Hyi and EK TrA.
Such modeling has been carried out previously for CV WDs 
such as WZ Sge \citep{sio95a} and AM Her \citep{gan06}, and also
for DA WDs \citep{koe96,koe98,dup03,dup06} and ZZ Ceti WDs
\citep{all04b}.  It is interesting to compare our results with 
these results as the presence/absence of the quasi-molecular hydrogen
lines in FUV spectra of WDs is not yet fully understood. Quasi-molecular 
satellite transitions take place during close collisions of the radiating 
hydrogen atom with a perturbing atom or proton. These absorption
features are present in the red wings of the Lyman series lines 
(Ly$\alpha$ in STIS and IUE spectral ranges at 1400 \& 1600\AA\ ; 
Ly$\beta$ at 1058 \& 1076\AA\ and Ly$\gamma$ around 995\AA\ - in 
the {\it{FUSE}} spectral range) and provide a noticeable source of opacity
in the hydrogen-rich atmosphere of     
WDs \citep{koe85,all94,all98,all04a,all04b,dup06}.  
In theory the H$_2^+$  satellite appearance is very sensitive to 
the degree of ionization and may be used as a temperature diagnostic. 
On the other hand, since the quasi-molecular opacity is proportional
to the proton and H\,{\sc i} densities, the quasi-molecular satellite
lines are expected to be observed at higher effective temperatures 
in WDs          with larger surface gravities \citep{koe96,dup03}. 
We summarize in Table 6 the occurrence of the quasi-molecular satellite 
lines in a number of WDs, covering a wide range of 
temperatures and gravities, listed in order of increasing temperature.  
We are aware that these WDs belong to different types (CVs, DA and ZZ Ceti)
and that the detection of the quasi-molecular satellite line also
depends on the S/N of the data. 
However, this table might help put restrictions on the upper and 
lower limits of the gravity and temperature of WDs, especially when 
the mass and distance are relatively unknown. For example it is clear
that the WD of EY Cyg must have $Log(g)<9.3$, otherwise one would
expect to see  the quasi-molecular satellite lines in the {\it{FUSE}} range.   
For WW Cet the opposite is true and one can therefore expect $Log(g)>7.8$.
In fact one could argue that for WW Cet $Log(g)>8.3$ while 
for EY Cyg $Log(g)<9.0$, as this would be even more consistent with 
Table 6 and with the errors on the WD masses of these systems.  
However, the system that really stands out in the list 
is AM Her which exhibits the 1076\AA\ feature but not the one at 1058\AA\ . 
\citet{gan06} suggest that the magnetic field plays a role in
the formation of these lines, and cite the fact the magnetic DA
WD PG 1658+441 has a weaker than expected 1058\AA\ 
$H_2^+$ absorption feature. However, \citet{dup03} mentioned that the
treatment of the quasi-molecular satellite opacities could be improved,
especially in the range of ionic densities encountered in the 
atmosphere of ultramassive WDs. The {\it{FUSE}} spectrum of PG 1658+441
actually shows that both the  1058\AA\ and 1076\AA\ absorption features
are rather weak, contrary to AM Her where the 1058\AA\ feature is
completely absent. We see the importance of generating a large database
of WDs exhibiting the quasi-molecular satellite lines of hydrogen, as it
could be used to assess the WD's temperature and/or gravity more accurately.

\section{Summary}

We have analyzed the FUV spectra of five DNs in quiescence with 
the aim of securing more accurate values of 
system parameters, such as the WD surface temperatures $T_{eff}$, 
their projected rotational velocities $V_{rot} \sin{i}$
(derived mainly from the {\it{FUSE}} data), their chemical
abundances, their surface gravities and the distances to the systems.
Four of these systems (EY CYG, SS Aur, VW Hyi,  and EK TrA) 
were part of our broader {\it{HST}} archival program with matching 
{\it{FUSE}} and {\it{HST}}/STIS spectra, while  
RU Peg has only one {\it{FUSE}} spectrum. 
We carried out the analysis including options for
the treatment of the hydrogen molecular satellite lines (for modeling 
cooler WDs) and NLTE atmosphere models (for modeling hotter WDs), 
dereddening the spectra and identifying the ISM molecular absorption 
lines in the {\it{FUSE}} spectra.
The following improvements (in methods and results) were achieved.

For EY Cyg, we disregarded the noisy portion of the {\it{FUSE}} spectrum
($\lambda < 950 $\AA ), 
which turned out to be 
an artifact of the re-processing and thus improved the fit. From the 
combined {\it{HST}}/STIS + {\it{FUSE}} spectrum 
we found $T_{eff} = 30,000$K and d=530pc assuming $Log(g)= 9$; and 
$T_{eff} = 32,000$K and d=310pc assuming $Log(g)= 9.5$.
We derived for the first time evidence of CNO processing in the 
photosphere of the WD itself in EY Cyg. 
Previously, only the emission line strengths implied suprasolar N and 
subsolar (deficient) C. Therefore EY Cyg's 
WD is only the third WD known to have anomalous N/C in its photosphere, 
the other objects being VW Hyi and U Gem. 

For SS Aur, we re-evaluated and improved the WD parameters by modeling 
for the first time for this object the combined {\it{FUSE}}+STIS spectra. 
We improved the fitting in the {\it{FUSE}} short wavelength range
without the need to introduce a second component, contrary to
\citep{sio04a}'s result where the second component  actually 
gave too much flux. This might be due to the fact that
\citet{sio04a} did not deredden the {\it{FUSE}} spectrum of SS Aur,
while we assumed E(B-V)=0.08. In addition this result is in excellent 
agreement with the distance  and we found no deficiency in carbon. 
We also improved the line identifications over the previously published 
{\it{FUSE}} analysis.  

For VW Hyi, the {\it{FUSE}} and STIS spectra were combined together for 
the first time. The opacity of the quasi-molecular satellite lines of 
hydrogen were included, and the spectrum was dereddened assuming E(B-V)=0.01.  
Our best model fit has $T_{eff}$=22,000K, $Log(g)=8.0$ and $d=60$pc.
Both the projected rotational velocity (400km/s) and the chemical abundances
are determined quite accurately due to the high S/N of the {\it{FUSE}} spectrum. 
We find a ratio C/N=0.25/3=1/12 and the silicon abundance is almost 
twice solar. A second component is detected in the very short wavelengths
contributing only a few percent of the total flux.
Even though we deliberately decided not to model the second component,
our results are in complete agreement with all the results from  
previous analyses of the system 
\citep{mat84,sio95a,gan96,sio96,sio97,sio01,god04}.   

For EK TrA we re-evaluated and improved the WD 
parameters by modeling for the first time the 
combined {\it{FUSE}}+STIS spectrum. We found a WD temperature of 
17,000K$\pm$1,000K assuming $Log(g)=8.0\pm 0.5$ 
leading to a distance of $\approx$125pc. \citet{gan01} carried
out a similar analysis, but found a somewhat higher 
temperature and a flux excess in the longer wavelengths, 
assuming a distance d$>$180pc.
The difference in results is probably due to 
our different modeling; namely we combined the {\it{FUSE}} and STIS
spectra, included the opacity of the quasi-molecular satellite lines
for hydrogen, and dereddened the spectrum.

For RU Peg we improved the line identifications over
the previously published {\it{FUSE}} analysis and we show that the
large mass of the WD gives a better agreement with a higher
temperature than previously estimated; namely the $Log(g)=8.8$
model gives a temperature of 70,000K for a distance of 282pc.  
We find that the projected rotational velocity is rather
small ($<$70km/s) and the carbon and silicon abundances 
are sub-solar; while sulfur and nitrogen are over-abundant. 
We assessed  the molecular hydrogen column density  
($N(H_2)=1\times 10^{16}$cm$^{-2}$), the atomic hydrogen column density
($N(HI) = 1\times 10^{17}-1\times 10^{18}$cm$^{-2}$),
and found that the lines are blue shifted with a velocity of -35km$~$s$^{-1}$. 

\acknowledgments

PG is thankful to Pierre Chayer for assistance in the computation 
of some of the synthetic spectra  and for a discussion on the some 
of the metal lines in the {\it{FUSE}} spectral range.
PG also wishes to thank Mario Livio, for his kind hospitality at the 
Space Telescope Science Institute, 
where part of this work was done, and 
Greg Masci (from the SDS Branch, ITS Division at STScI) 
for a computer upgrade and additional data disks.  
Support for this work was provided by NASA through
grant numbers HST-AR-10657.01-A (HST Cycle 14 Archival,
to Villanova University, P.Godon) and GO-09724.06A (to the
University of Washington, P.Szkody)  
from the Space Telescope Science Institute,
which is operated by the Association of Universities for
Research in Astronomy, Incorporated, under NASA contract NAS5-26555.
This work was also supported in part by NSF grant AST0507514
and NASA grant NNG04GE78G, both to Villanova University (E.M. Sion). 
This research was partly based on observations made with the
NASA-CNES-CSA Far Ultraviolet Spectroscopic Explorer.
{\it{FUSE}} is operated for NASA by the Johns Hopkins University under
NASA contract NAS5-32985.

\clearpage 

\setlength{\hoffset}{-10mm} 
\begin{table} 
\caption{Parameters of the Systems}
\begin{tabular}{lllllllllll}
\hline
Name     &Type &$E_{B-V}$& d  & P$_{orb}$& $i$     &Spectral&$M_{1} $      &$M_{2}$      &V$_{max}$&V$_{min}$\\
         &     &       & (pc) & (hrs)    & (deg)   &  Type  &$(M_{\odot})$   &$(M_{\odot}) $ &       &       \\
\hline
EY Cyg   &DN UG& 0.00 & 450   & 11.02377 &$16\pm 1$& K7 V   & $1.26 \pm 0.29$& $0.59\pm0.13$ & 11.4  & 15.5  \\
SS Aur   &DN UG& 0.08 & 200   & 4.3872   &$38\pm16$& M1 V   & $1.08 \pm 0.40$& $0.39\pm0.02$ & 10.5  & 14.5  \\
VW Hyi   &DN SU& 0.01 &  65   & 1.783    &$60\pm10$& L0     & $0.63-0.86    $& $0.11\pm0.02$ &  8.5  & 13.8  \\
EK TrA   &DN SU& 0.03 &  180  & 1.5091   &$58\pm7 $&        & $0.46 \pm 0.10$& $0.09\pm0.02$ &  12.0 & 17.0  \\
RU Peg   &DN UG& 0.00 & 282   & 8.9904   &$33\pm5 $& K2-5V  & $1.29 \pm 0.20$& $0.94\pm0.04$ &   9.0 & 13.1  \\ 
\hline
\end{tabular}
{\small References.
EY Cyg: \citet{sar95,smi97,cos98,tov02,cos04,sio04b};  
SS Aur: \citet{sha83,sha86,har99};  
VW Hyi: \citet{sch81,van87a,van87b,sio95b,hua96b,lon96,whe96,sio97};   
EK TrA: \citet{vog80,war87,gan97,gan01};
RU Peg: \citet{sto81,wad82,sha83,fri90,joh03}.  
The E(B-V) values are from \citet{ver87,lad91,bru94}.  
}
\end{table} 

\clearpage 

\begin{table} 
\caption{FUV Observations of the Systems: {\it{HST}} \& {\it{FUSE}} Spectra} 
\begin{tabular}{lclcllll}
\hline
System & Date      & Telescope   & Exp time & Dataset   & Filter/Grating &Operation& $\tau_q$\tablenotemark{a}   \\ 
       &(dd/mm/yy) & /Instrument & (ksec)   &           & /Aperture      & Mode & (days)    \\ 
\hline
EY Cyg   & 16-07-03  & FUSE        & 18       & D1450101  & LWRS           & TTAG  & dq\tablenotemark{b} \\
         & 10-02-03  & STIS        & 0.7      & O6LI0V010 & G140L/52X0.2   & ACCUM & dq\tablenotemark{b} \\
SS Aur   & 13-02-02  & FUSE        & 14.5     & C1100201  & LWRS           & TTAG  & 28  \\
         & 20-03-03  & STIS        & 0.6      & O6LI0F010 & G140L/52X0.2   & ACCUM & 30  \\
VW Hyi   & 18-08-01  & FUSE        & 18       & B0700201  & LWRS           & TTAG  & 11   \\
         & 01-06-00  & STIS        & 2.5      & O5E205010 & E140M/0.2X0.2  & ACCUM & 1  \\
         & 06-06-00  & STIS        & 2.5      & O5E206010 & E140M/0.2X0.2  & ACCUM & 7  \\
         & 10-12-01  & STIS        & 9.1      & O5E201010 & E140M/0.2X0.2  & ACCUM & 15  \\
EK TrA   & 24-06-02  & FUSE        & 33       & Z9104301  & LWRS           & TTAG  & 45 \\
         & 24-06-02  & FUSE        & 33       & Z9104302  & LWRS           & TTAG  & 45 \\
         & 25-07-99  & STIS        & 4.3      & O5B612010 & E140M/0.2X0.2  & TTAG  & 155 \\ 
RU Peg   & 05-07-02  & FUSE        & 2.9      & C1100101  & LWRS           & TTAG  & 60 \\ 
\hline
\end{tabular}
\tablenotetext{a} {in the last column $\tau_q$ is the time in quiescence, i.e. since outburst} 
\tablenotetext{b} {for the very long recurrence time systems (EY Cyg \& BZ UMa), the
observations were made in deep quiescence (dq)}
\end{table} 

\clearpage 

\begin{table} 
\caption{{\it{FUSE}} Lines} 
\begin{tabular}{llllllll}
\hline
Line           & Wavelength   & Origin  & EY Cyg & SS Aur & VW Hyi & EK TrA & RU Peg  \\
Identification & (\AA )       &         &        &        &        &        &         \\
\hline
H\,{\sc i}    &               & c,ism   &  e,c   &  e,c   &  e,c   &  e,c   &   e,c   \\
O\,{\sc i}    &               & c,ism   &  e,c   &   -    &   -    &  e,c   &    -    \\
N\,{\sc iv}   & 921.5-924.9   &    s    &  e,c   &   -    &   -    &   -    &    a    \\
S\,{\sc vi}   & 933.5         &    s    &   -    &   -    &   -    &   -    &    a    \\
S\,{\sc vi}   & 944.5         &    s    &   -    &   -    &   -    &   -    &    a    \\
C\,{\sc iii}  & 977.0         &    s    &   -    &   e    &   e    &   e    &    e    \\
N\,{\sc iii}  & 979.8         &    s    &   -    &   -    &   -    &   -    &    a    \\
N\,{\sc iii}  & 991.6         &    s    &   c    &   c    &   -    &   -    &    c    \\
C\,{\sc ii}   & 1010.         &    s    &   -    &   a    &   -    &   -    &    -    \\
Si\,{\sc ii}  & 1020.7        &   ism   &   -    &   a    &   -    &   -    &    -    \\
O\,{\sc vi}   & 1031.9        &   s     &   e    &   e    &   e    &   e    &   a,e   \\
              & 1037.6        &   s     &   e    &   c    &   e    &   e    &   a,e   \\
C\,{\sc ii}   & 1036.3        &   ism   &   a    &   -    &   a    &   a    &    -    \\
S\,{\sc iv}   & 1062.6        &    s    &   -    &   c    &   -    &   -    &    a    \\
C\,{\sc ii}   & 1066.        &    s    &   -    &   a    &   a    &   -    &    -    \\
Si\,{\sc iv}  & 1066.6        &    s    &   -    &   -    &   -    &   -    &    a    \\
O\,{\sc iv}   & 1067.8        &    s    &   -    &   -    &   -    &   -    &    a    \\
S\,{\sc iv}   & 1073.0,1073.5 &    s    &   a    &   a    &   -    &   -    &    a    \\
N\,{\sc ii}   & 1084.0-1085.7 & s,c,ism &   c    &   c    &   a    &   c    &    c    \\
Si\,{\sc iii} & 1108.4-1113.2 &   s     &   -    &   -    &   a    &   a    &    -    \\
C\,{\sc iii}  & 1125.6        &   s     &   a    &   -    &   -    &   a    &    -    \\
C\,{\sc ii}   & 1127          &   s     &   -    &   -    &   -    &   a    &    -    \\
Si\,{\sc iv}  & 1122.5        &   s     &   a    &   a    &   a    &   a    &    a    \\
              & 1128.3        &   s     &   a    &   a    &   a    &   a    &    a    \\
N\,{\sc i}    & 1134.2-1135.0 & s,c,ism &   c    &   c    &   c    &   c    &    c    \\
Si\,{\sc iii} & 1140.5-1145.7 &   s     &   a    &   a    &   a    &   -    &    -    \\
Si\,{\sc iii} & 1158.1        &   s     &   -    &   -    &   -    &   a    &    -    \\
He\,{\sc ii}  & 1168.6        &   c     &  e,c   &   -    &  e,c   &  e,c   &   e,c   \\
C\,{\sc iii}  & 1174.9-1176.4 &   s     &   -    &   -    &   a    &  e,a   &   e,a   \\
\hline
\end{tabular}
\tablenotetext{}{The following abbreviations have been used: a - for absorption; 
e - for emission; c - for contamination (e.g. air glow); s - for source; and
ism - for interstellar medium. 
}
\end{table}

\clearpage 

\begin{table} 
\caption{STIS Lines} 
\begin{tabular}{lllllll}
\hline
Line           & Wavelength   & Origin  & EY Cyg &  SS Aur & VW Hyi & EK TrA   \\
Identification & (\AA )       &         &        &         &        &          \\
\hline
C\,{\sc iii}  & 1174.9-1176.4 &         &   -    &    -    &   a    &    e     \\
Si\,{\sc ii}  & 1190.4        &         &  a,e?  &    -    &   -    &    e?    \\
              & 1193.3        &         &  a,e?  &    -    &   -    &    e?    \\
              & 1194.5        &         &  a,e?  &    -    &   -    &    e?    \\
Si\,{\sc iii} & 1206.5        &         &   e    &    e    &   -    &    e     \\
H\,{\sc i}    & 1215.7        &         &   a,c  &    a    &  a,e,c &    a,c   \\
N\,{\sc v}    & 1238.8        &         &   e    &    e    &    -   &    e     \\
              & 1242.8        &         &   e    &    e    &    -   &    e     \\
Si\,{\sc ii}  & 1259.9, 1260.4&         &   -    &    a    &    a   &    a     \\
              & 1264.7, 1265.0&         &   -    &    a?   &    a   &    a     \\
C\,{\sc i}    & 1266.4        &         &   a?   &    -    &    a   &    -     \\
Si\,{\sc iii} & 1298.9        &         &   a    &    a    &    a   &    a?    \\
Si\,{\sc ii}  & 1304.4        &         &   a    &    -    &    a   &    a?    \\
N\,{\sc i}    & 1316.0        &         &   -    &    -    &    -   &    -     \\
              & 1319.7        &         &   -    &    -    &    -   &    -     \\
C\,{\sc i}    & 1328.8, 1329.6&         &   -    &    -    &    a   &    -     \\
C\,{\sc ii}   & 1334.5, 1335.7&         &   a    &   a,e?  &    a   &    -     \\
C\,{\sc i}+   & $\approx$1350 &         &   -    &    -    &    a   &    a,e   \\
Si\,{\sc iv}  & 1393.8        &         &   e    &   a,e?  &   e,a? &    e     \\
              & 1402.8        &         &   e    &   a,e?  &   e,a? &    e     \\
Ni\,{\sc ii}  & 1467.3, 1467.7&         &   -    &    -    &    -   &    -     \\
S\,{\sc i}    & 1485.6, 1487.2&         &   -    &    -    &    a   &    -     \\
N\,{\sc i}    & 1492.6, 1492.8&         &   -    &    -    &    a   &    -     \\
              & 1494.7        &         &   -    &    -    &    a   &    -     \\
Si\,{\sc ii}  & 1526.7        &         &   -    &    -    &    a   &    a     \\
              & 1533.4        &         &   -    &    -    &    a   &    a     \\
C\,{\sc iv}   & 1548.2        &         &   -    &    e    &    e   &    e     \\
              & 1550.8        &         &   -    &    e    &    e   &    e     \\
C\,{\sc i}    & 1560.3-1561.4 &         &   -    &    -    &    a   &    -     \\
He\,{\sc ii}  & 1640.5        &         &   e    &    -    &    -   &    e     \\
C\,{\sc i}    & 1656.3-1658.1 &         &   -    &    -    &    a?  &    a     \\
Al\,{\sc ii}  & 1670.8        &         &   -    &    -    &    a?  &    a     \\
\hline
\end{tabular}
\tablenotetext{}{The following abbreviations have been used: a - for absorption; 
e - for emission; c - for contamination (e.g. air glow); s - for source; and
ism - for interstellar medium. 
}
\end{table}

\clearpage 

\setlength{\hoffset}{-23mm} 
\begin{table} 
\caption{Synthetic Stellar Spectra} 
\begin{tabular}{ccccccccrccc}
\hline
Name & $Log(g)$ & T &$V_{rot} sin(i)$& [C]   & [Si]  & [N]   & [S] & d &f\tablenotemark{a}& $\chi^2_{\nu}$ &Figure \\ 
       &(cm$~$s$^{-2}$)&($10^3$K)&(km$~$s$^{-1}$)   & Solar & Solar & Solar & Solar &(pc) &              &                &       \\ 
\hline
EY Cyg &  9.00    &30.0$\pm 2.0$&100$\pm$50&0.03$\pm$0.02& 0.6$\pm0.1$ & 2.0 $\pm1.0$ & 6.0$\pm1.0$ &530$\pm 60$&1.50&1.140&1,2   \\
       &  9.50    &32.0$\pm 2.0$&100$\pm$50&0.03$\pm$0.02& 0.6$\pm0.1$ & 2.0 $\pm1.0$ & 6.0$\pm1.0$ &310$\pm 40$&1.50&1.046& -     \\ 
SS Aur &  8.31    &27.0$\pm 1.0$&400$\pm$100     & 1.0   & 1.0         & 1.0          & 1.0         & 200 & 0.93 & 1.476  & -     \\
       &  8.31    &30.0$\pm 1.0$&400$\pm$100     & 1.0   & 1.0         & 1.0          & 1.0         & 254 & 0.93 & 1.472  & -     \\
       &  8.71    &31.0$\pm 1.0$&400$\pm$100     & 1.0   & 1.0         & 1.0          & 1.0         & 200 & 0.93 & 1.474  & -     \\
       &  8.71    &33.0$\pm 1.0$&400$\pm$100     & 1.0   & 1.0         & 1.0          & 1.0         & 240 & 0.93 & 1.472  & -     \\
       &  8.93    &34.0$\pm 1.0$&400$\pm$100     & 1.0   & 1.0         & 1.0          & 1.0         & 200 & 0.93 & 1.471  & 3,4   \\
VW Hyi &  8.00    &22.0$\pm 1.0$&400$\pm$100 &0.25$\pm$0.05&1.8$\pm$0.2&3.0$\pm1.0$   & 1.0         & 60  & 1.00 & 0.734  & 5,6   \\ 
       &  8.50    &24.0$\pm 1.0$&400$\pm$100 &0.25$\pm$0.05&1.8$\pm$0.2&3.0$\pm1.0$   & 1.0         & 51  & 1.00 & 0.770  & -     \\
EK TrA &  7.50    &15.5$\pm 0.5$&200$\pm$100 & 0.1$\pm0.05$&0.6$\pm0.1$& 1.0          & 1.0         & 137 & 1.00 & 0.781  & -     \\
       &  8.00    &17.0$\pm 0.5$&200$\pm$100 & 0.1$\pm0.05$&0.6$\pm0.1$& 1.0          & 1.0         & 126 & 1.00 & 0.781  & 7,8   \\
       &  8.50    &18.0$\pm 0.5$&200$\pm$100 & 0.1$\pm0.05$&0.6$\pm0.1$& 1.0          & 1.0         & 104 & 1.00 & 0.780  & -     \\
RU Peg &  8.55    &50.0$\pm2.0$&40/-15+30&0.2$\pm0.1$&0.2$\pm0.1$      &$\ge$1.0      &10$\pm5$     & 282 &  --  & 0.565  & -     \\ 
       &  8.72    &60.0$\pm5.0$&40/-15+30&0.2$\pm0.1$&0.2$\pm0.1$      &$\ge$1.0      &10$\pm5$     & 282 &  --  & 0.545  & -     \\ 
       &  8.80    &70.0$\pm5.0$&40/-15+30&0.2$\pm0.1$&0.2$\pm0.1$      &$\ge$1.0      &10$\pm5$     & 282 &  --  & 0.539  & 10,11 \\ 
       &  8.92    &80.0$\pm5.0$&40/-15+30&0.2$\pm0.1$&0.2$\pm0.1$      &$\ge$1.0      &10$\pm5$     & 282 &  --  & 0.553  & -     \\ 
       &  9.23   &110.0$\pm5.0$&40/-15+30&0.2$\pm0.1$&0.2$\pm0.1$      &$\ge$1.0      &10$\pm5$     & 282 &  --  & 0.827  & -     \\ 
\hline
\end{tabular}
\tablenotetext{a}{
The factor by which the STIS spectrum had to be scaled so that its flux level (continuum) 
matches the flux level (continuum) of the corresponding {\it{FUSE}} spectrum at the overlapping 
wavelengths $\approx 1150-1185$\AA }.  

\end{table}

\clearpage

\setlength{\hoffset}{-00mm} 
\begin{table} 
\caption{Presence of Quasi-Molecular Satellite Lines ($H_2^+$) in White Dwarfs} 
\begin{tabular}{llllcc}
\hline
Name       & type &  $T_{eff}$ & $Log(g)$ &  FUSE range  & STIS range \\
           &         &(1000K)  &        & 1058,1076\AA &  1,400\AA \\
\hline
G226-29    & ZZ Ceti & 12.0    & 7.93   &  yes, yes    &   -     \\ 
G185-32    & ZZ Ceti & 12.0    & 7.9    &  yes, yes    &   -     \\ 
WZ Sge     & DN UG   & 14.0    & 8.5    &      -       &   yes   \\
WZ Sge     & DN UG   & 15.5    & 8.5    &      -       &   weak  \\
40 Eri B   & DA WD   & 16.5    & 7.77   &   yes, yes   &   -     \\ 
EK TrA     & DN SU   & 17.0    & 8.0    &   yes, yes   &   weak  \\  
BZ UMa     & DN SU   & 17.5    & 8.5    &      -       &   yes   \\ 
AM Her     & NL AM   & 19.8    & 8.2    &    no, yes   & very weak \\ 
Wolf 1346  & DA WD   & 19.9    & 7.84   &   yes, yes   &   -     \\ 
He 3       & DA WD   & 21.4    & 8.04   &   yes, yes   &   -     \\ 
GD 140     & DA WD   & 21.6    & 8.41   &   yes, yes   &   -     \\ 
VW Hyi     & DN SU   & 22.0    & 8.0    &   yes, yes   &   no    \\
WZ Sge     & DN UG   & 23.2    & 8.5    &   yes, yes   &   -     \\ 
L825-14    & DA WD   & 25.0    & 7.76   &    no, no    &   -     \\ 
WW Cet     & DN UG   & 26.0    & 8.3    &   yes, yes   &   no    \\
PG 1658+441&DA WD    & 29.6    & 9.31   &  weak, yes   &   -     \\ 
EY Cyg     & DN UG   & 30.0    & 9.0    &    no, no    &   no    \\
\hline
\end{tabular}
\tablenotetext{}{References. 
G226-29 \& G185-32: \citet{all04b}; 
BZ UMa: based on an inspection of the STIS spectrum compared to 
a synthetic spectrum;  
EK TrA: this work; 
AM Her: \citet{gan06}; VW Hyi: this work; WW Cet: the quasi-molecular
feature was detected but not modeled in \citet{god06b};
PG 1685+441: \citet{dup03,dup06}; all the other DA WDs are from
the ORFEUS observations of \citet{koe98} . 
WZ Sge: all the modelings carried out in \citet{sio95a,god06a,lon03}
assumed a 0.9 solar mass. 
The 2004 July STIS spectrum of WZ Sge (with $T_{eff}$=15,500K) 
was analyzed again in the present work to detect the quasi-molecular 
hydrogen feature around 1400\AA\ : it was found to be rather weak.}     
\end{table}

\clearpage

FIGURE CAPTIONS

\figcaption{ \\ 
The combined {\it{FUSE}}+STIS spectrum of EY Cyg (light grey/red line) is
shown together 
with a 30,000K single WD model (thick black line) assuming $Log(g)=9.0$. 
The regions
masked during fitting are shown in blue (dark grey). 
With a flux of only a few $10^{-15}$ergs$~$s$^{-1}$cm$^{-2}$\AA$^{-1}$,
EY Cyg is a relatively weak source contaminated with many air glow
emission lines in the {\it{FUSE}} range, especially in the shorter wavelength.
The increase of flux at $\lambda < 934$\AA\ is 
masked in the modeling (see text).  
The WD model has a projected rotational velocity 
of 100km$~$s$^{-1}$, a very low carbon abundance (note the absence
of the C\,{\sc iii} (977\AA ) and C\,{\sc iv} (1550\AA ) emission lines )
and $\chi^2_{\nu}=1.108$. 
The distance obtained from fitting the fluxes is $\approx 500$pc.  
}

\figcaption{ 
The 30,000K WD model (from Fig.1) is shown here with the 
{\it{FUSE}} spectra of EY Cyg (binned here 
at 0.1\AA\ ) and line identifications. 
Air glow lines are annotated with  a plus sign inside a circle; 
the O\,{\sc i}, N\,{\sc i} and N\,{\sc ii}
emission lines are also due to air glow; 
the interstellar H$_2$ molecular lines have been labeled vertically; 
the O\,{\sc vi} doublet emission is from the source; 
-- all these emission lines 
together with the ISM molecular hydrogen absorption lines have 
been masked in the fitting.  
The abundance of C, S, N and Si have been set to 0.03, 6.0, 2.0 and 
0.6 solar (respectively) by fitting the following absorption lines: 
S\,{\sc iv} (1173\AA ),  C\,{\sc ii} (1010\AA, 1066\AA ), 
C\,{\sc iii} (1175\AA ), Si\,{\sc iii} (1140-1145\AA ) and 
N\,{\sc ii} (1184-1186 \AA ). 
}

\figcaption{ 
The combined {\it{FUSE}}+STIS spectrum 
of SS Aur (in red/light grey) together with the best fit WD model
(in black). 
The regions that have been masked for the fitting are shown in 
blue (dark grey). 
The synthetic stellar spectrum has a temperature of 34,000K (assuming
$Log(g)=8.93$ and a distance of 200pc), a projected rotational 
velocity of 400km$~$s$^{-1}$ and solar abundances. 
The spectrum has been dereddened assuming E(B-V)=0.08. 
}

\figcaption{ 
The {\it{FUSE}} spectrum of SS Aur (binned here at 0.1 \AA\ ) 
is shown with the best fit model shown in Figure 3. Note that
the C\,{\sc iii} (977 \AA) emission indicates that there could possibly
be some emission at 1175\AA , which would make it difficult to detect 
absorption there, potentially explaining the observed flat spectrum in
that region. For that reason the carbon abundance was fitted using the  
C\,{\sc ii} lines seen in absorption both at 1010 \AA\ and 1066 \AA .
The sharp emission lines due to air glow and the ISM molecular hydrogen  
absorption lines have been masked for the fitting. 
}
 
\figcaption{ 
The combined {\it{FUSE}}+STIS spectrum of VW Hyi is shown in red
(light grey) together with the best fit WD model (in black). 
The emission lines that have been masked are shown in blue (dark grey). 
The synthetic stellar spectrum has a temperature
$T=22,000K$ (assuming $Log(g)=8.0$), a projected rotational velocity
of $V_{rot} sin(i)=400$km$~$s$^{-1}$, and non-solar composition: 
C=$0.25\times$solar, Si=$1.8\times$solar, N=$3.0\times$solar. 
The spectrum has been dereddened assuming
E(B-V)=0.01 and the distance obtained from the fitting is 60pc. 
}

\figcaption{
The {\it{FUSE}} portion of Figure 6 is shown with line identifications.
The dotted-dashed line shows the same model without the inclusion
of the quasi-molecular satellite lines at 1058\AA\  and 1076\AA\
(shown with arrows). The synthetic spectrum does not attempt
to model the O\,{\sc vi} doublet and C\,{\sc iii} (977\AA ) emission
lines and the low-level flux flat continuum in the lower wavelengths.
}

\figcaption{ 
The combined {\it{FUSE}}+STIS spectrum of EK TrA is shown in red (light 
grey) together with the best fit WD model (in black). 
The strong emission lines that have been masked for the fitting 
are shown in blue (dark grey). 
The synthetic stellar spectrum has a temperature
$T=17,000$K (assuming $Log(g)=8.0$), a projected rotational velocity 
of 200km$~$s$^{-1}$, and the following abundances (in solar units):
C=0.1 and Si=0.6. The spectrum was dereddened assuming E(B-V)=0.03
and the distance obtained from the fitting is d=126pc. 
}

\figcaption{ 
The {\it{FUSE}} spectrum of EK Tra (in red/light grey) is shown
with the best fit model presented in Figure 6 and line identifications. 
The dotted-dashed line shows the same model without the inclusion
of the quasi-molecular satellite lines modeling, at 1058\AA\  and 1076\AA\ . 
With a flux of 
$\approx 1 \times 10^{-14}$ergs$~$s$^{-1}$cm$^{-2}$\AA$^{-1}$
and less, EK TrA is actually a weak {\it{FUSE}} source. 
All the sharp emission lines are due to contamination (helio- and geo-coronal
emission) and have been masked for the fitting. 
The broad C\,{\sc iii} (977\AA\ \& 1175\AA\ ) 
and O\,{\sc vi} ($\sim 1130-1140$\AA\ ) emission lines are associated
with the source.  The synthetic spectrum does not attempt
to model the O\,{\sc vi} doublet and C\,{\sc iii} emission
lines and the low-level flux flat continuum in the lower wavelengths 
(which is of the level of the noise).   
The C\,{\sc ii} (1036\AA) absorption line is interstellar. 
}

\figcaption{ 
The {\it{FUSE}} spectrum of RU Peg in quiescence.
The sharp ISM molecular hydrogen absorption lines 
have been annotated as in Figure 2.  
The very broad emission lines of the 
O{\sc vi} doublet and 
C{\sc iii} (977 \& 1175 \AA ) are clearly 
apparent. In addition there are absorption lines
which originate only at very high temperatures ($T \ge 50,000$K)
such as
N\,{\sc iv} (around 923\AA ),  
S\,{\sc vi} (933.5 \& 944.5\AA ),  
and O\,{\sc vi} (1031.7 \& 1037.3\AA ).  
The additional sharp absorption lines (from low ionization levels such as 
N\,{\sc i}) are either from the ISM or due to air contamination
(or both).  
Though the flux is $\approx 1 \times 10^{-13}$ergs$~$s$^{-1}$cm$^{-2}$
\AA$^{-1}$, the spectrum is clearly under-exposed (1ksec) and has a low
S/N. 
}

\figcaption{ 
The best fit WD model to the {\it{FUSE}} spectrum of RU Peg is shown in this
Figure. 
The synthetic spectrum is in solid black, the
observed spectrum is in red (light grey), 
the regions of the observed spectrum
masked for the fitting are shown in blue (dark grey). 
The synthetic stellar spectrum
consists of 70,000K WD, with a projected rotational 
velocity of 40km$~$s$^{-1}$, 
carbon abundance of 0.2 times solar, silicon abundance of 0.2 times solar, 
and sulfur abundance of 10 times solar.  The absorption lines that have been
used to produce the abundance fits are annotated. 
}


\figcaption{ 
A detail of the fitting of the absorption lines around 1070\AA\
(note the Y-axis scale!).  
The molecular hydrogen lines have been marked below the spectrum
for clarity. 
The absorption lines shown are 
S\,{\sc iv} (1062.7\AA),  
Si\,{\sc iv} (1066.6\AA),  
O\,{\sc iv} (1067.8\AA), and the  
S\,{\sc iv} doublet (around 1073.5\AA).
The ISM model does not include metals and therefore the 
Ar\,{\sc i} and Fe\,{\sc ii} lines are not modeled, though they
are present in the observed spectrum. 
Note the redshift of the S\,{\sc iv} doublet.  
}
 
\clearpage 

\begin{figure}
\plotone{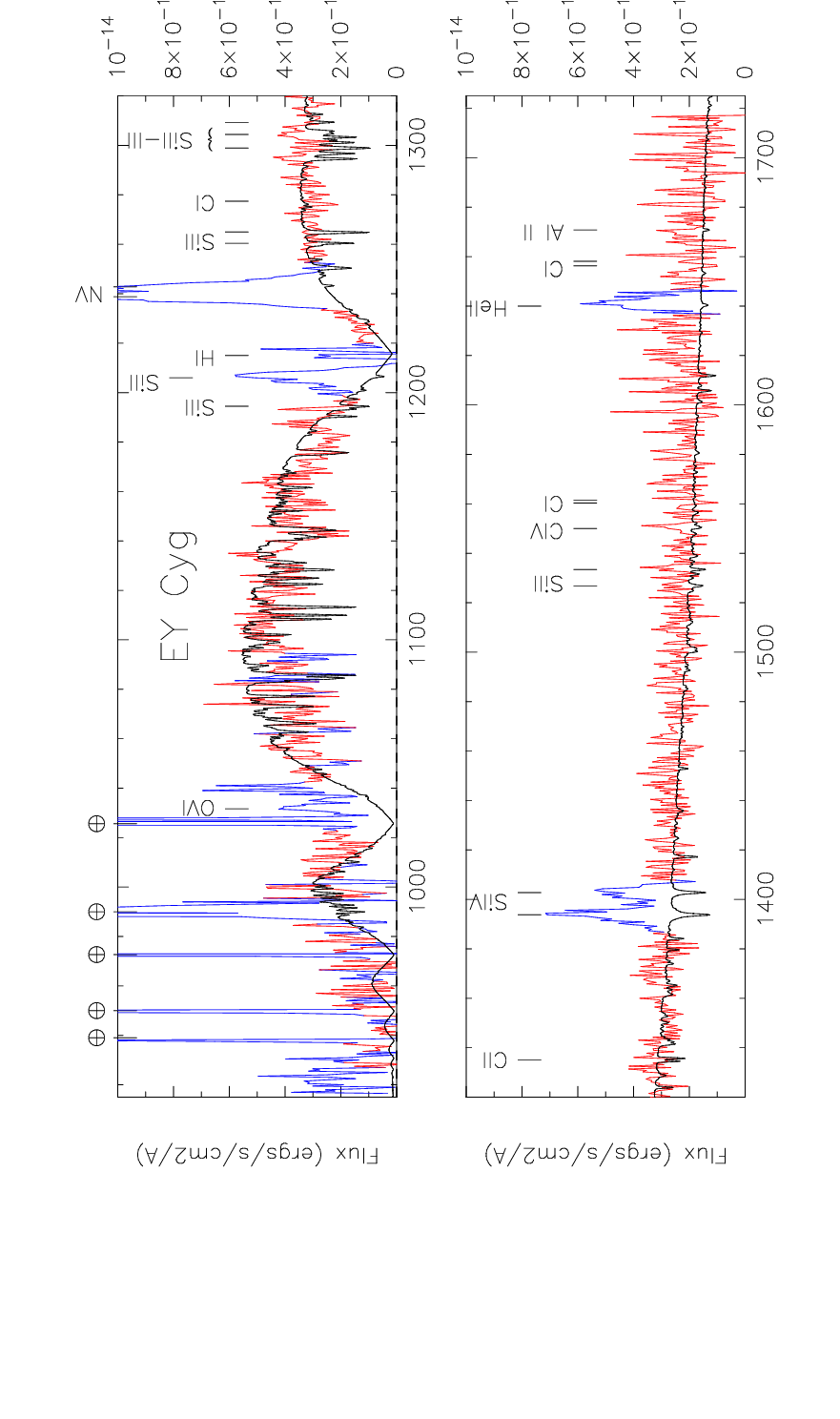}             
\end{figure}

\clearpage 

\begin{figure}
\plotone{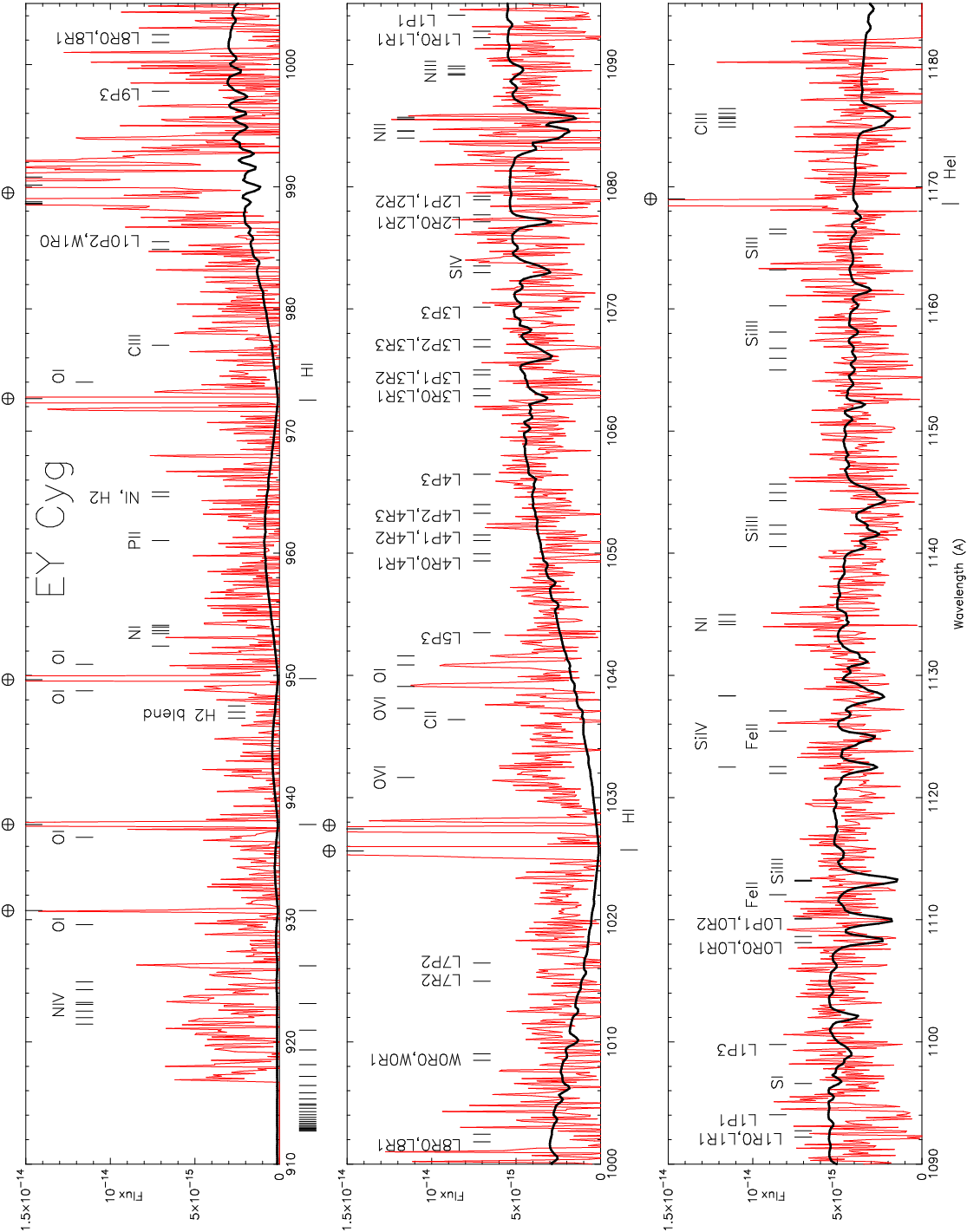}             
\end{figure} 
 
\clearpage

\begin{figure}
\plotone{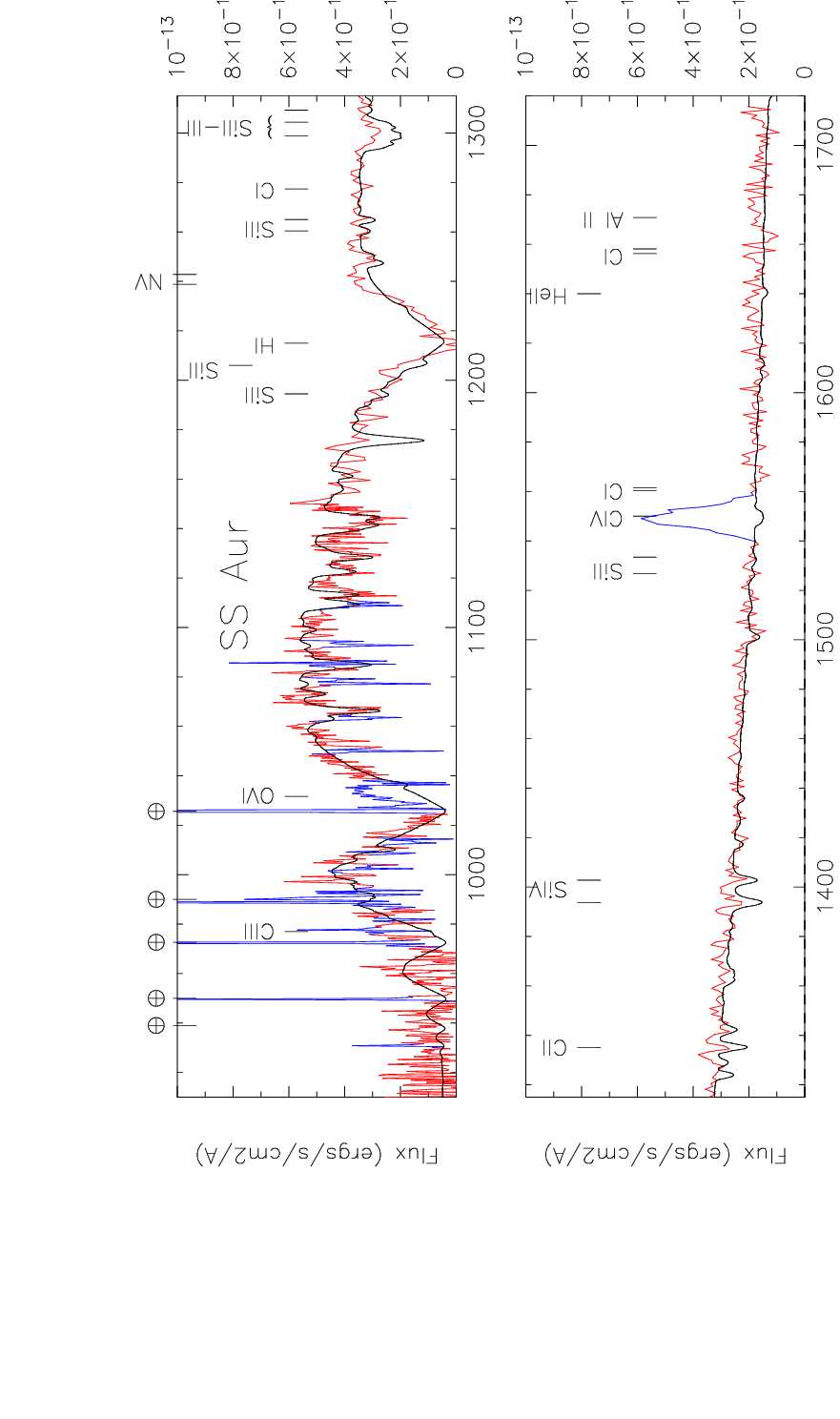}                         
\end{figure} 

\clearpage 

\begin{figure}
\plotone{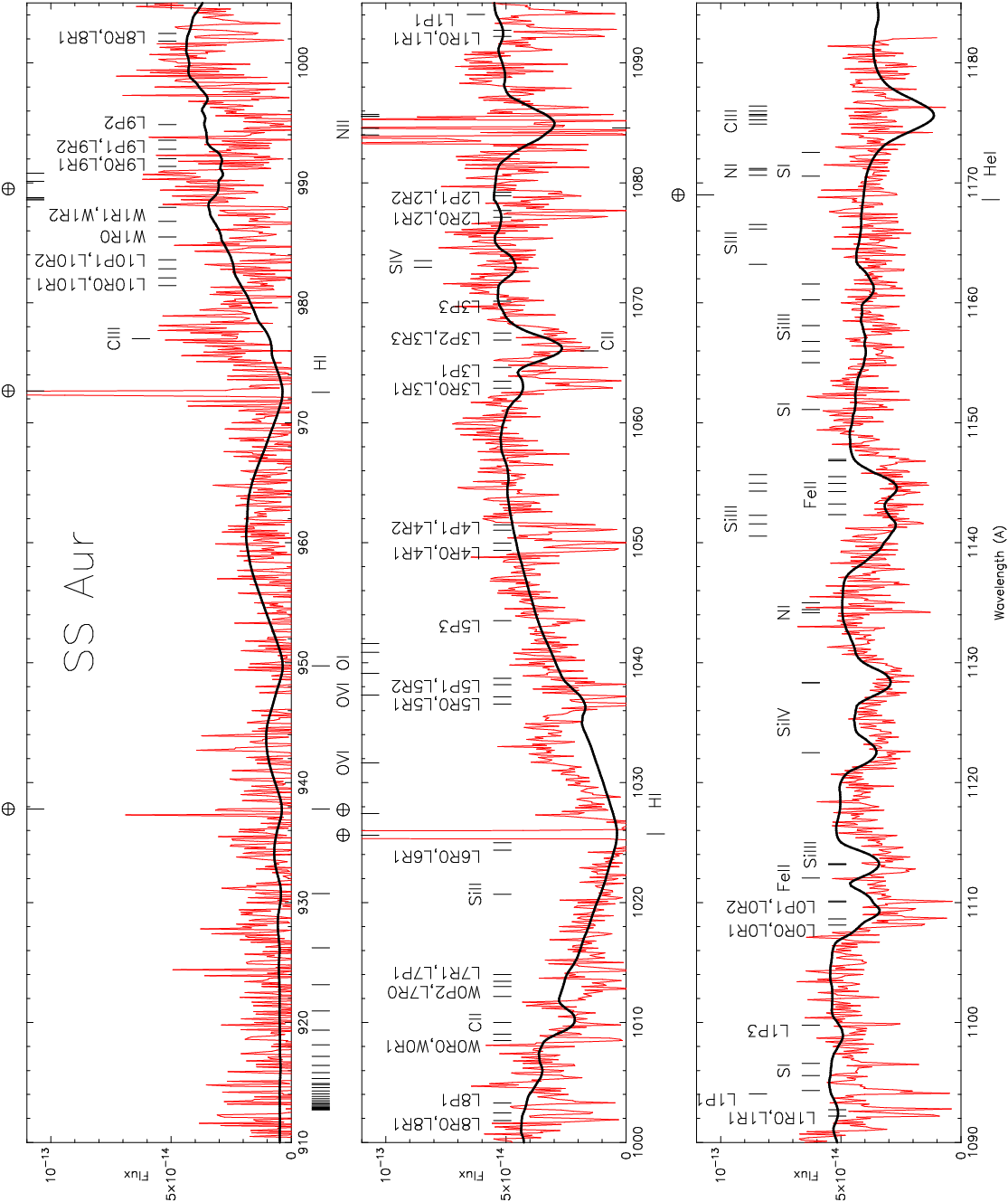}                                     
\end{figure} 

\clearpage 

\begin{figure}
\plotone{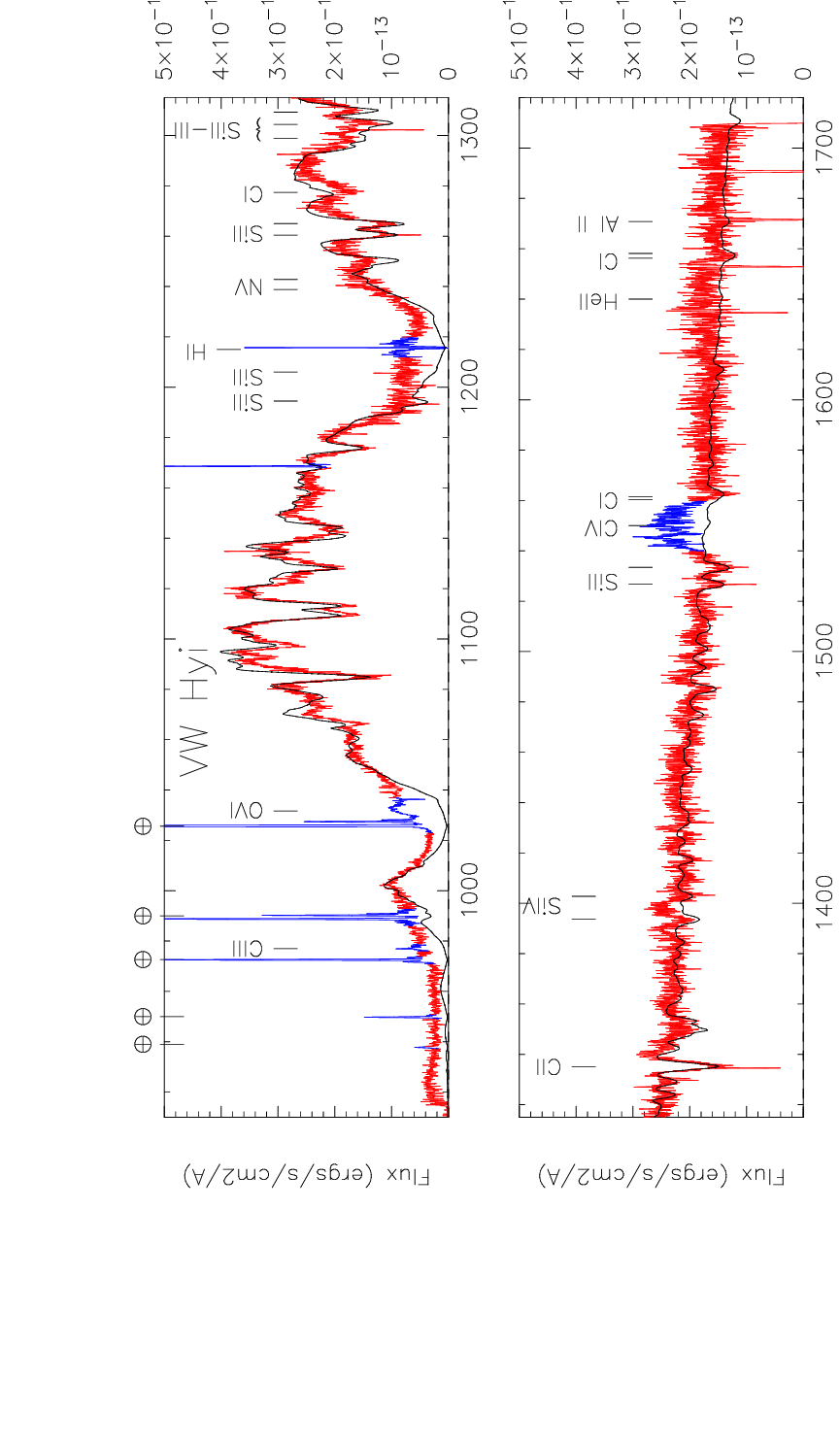}                                     
\end{figure}

\clearpage 

\begin{figure}
\plotone{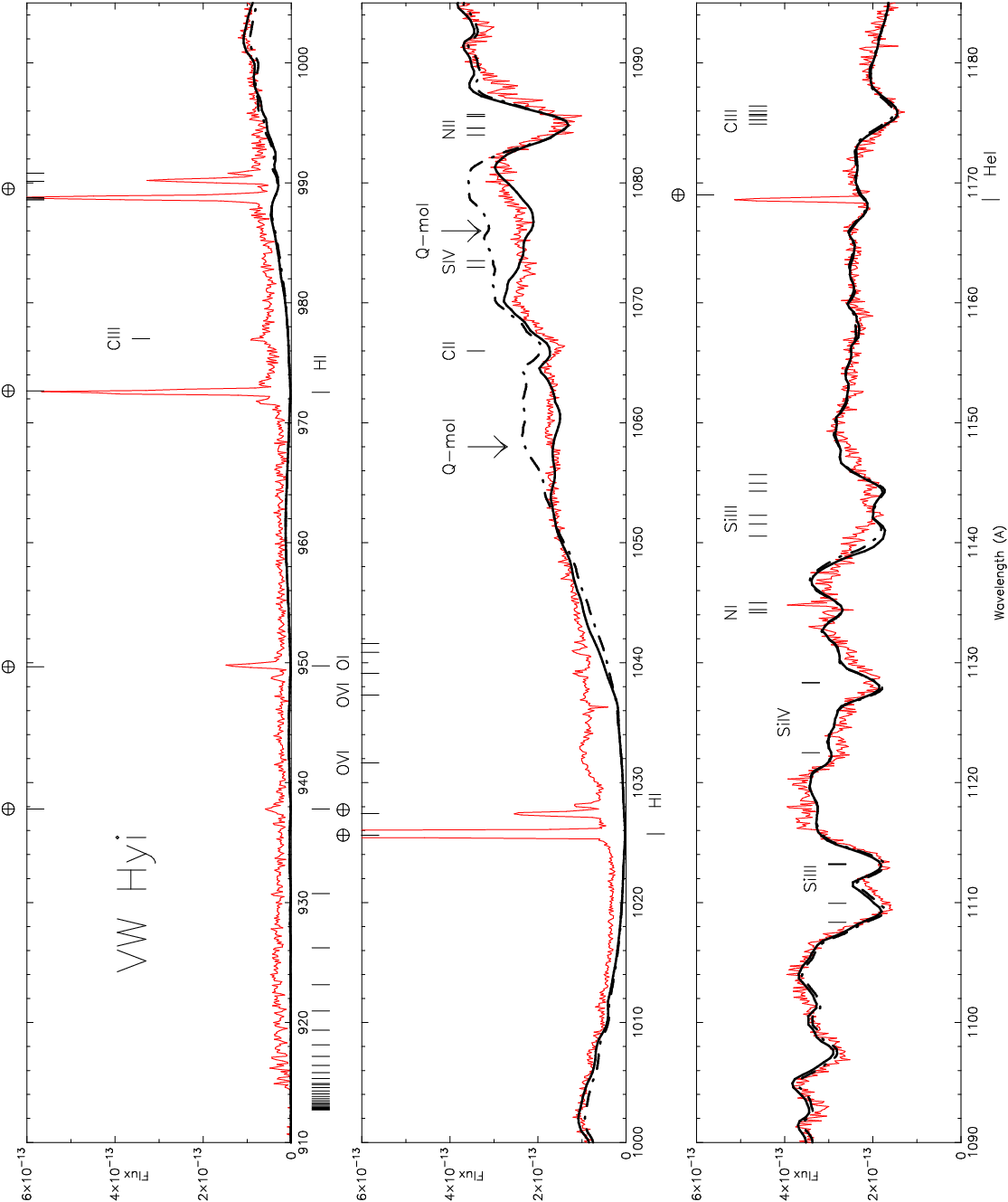}                                     
\end{figure} 

\clearpage 

\begin{figure}
\plotone{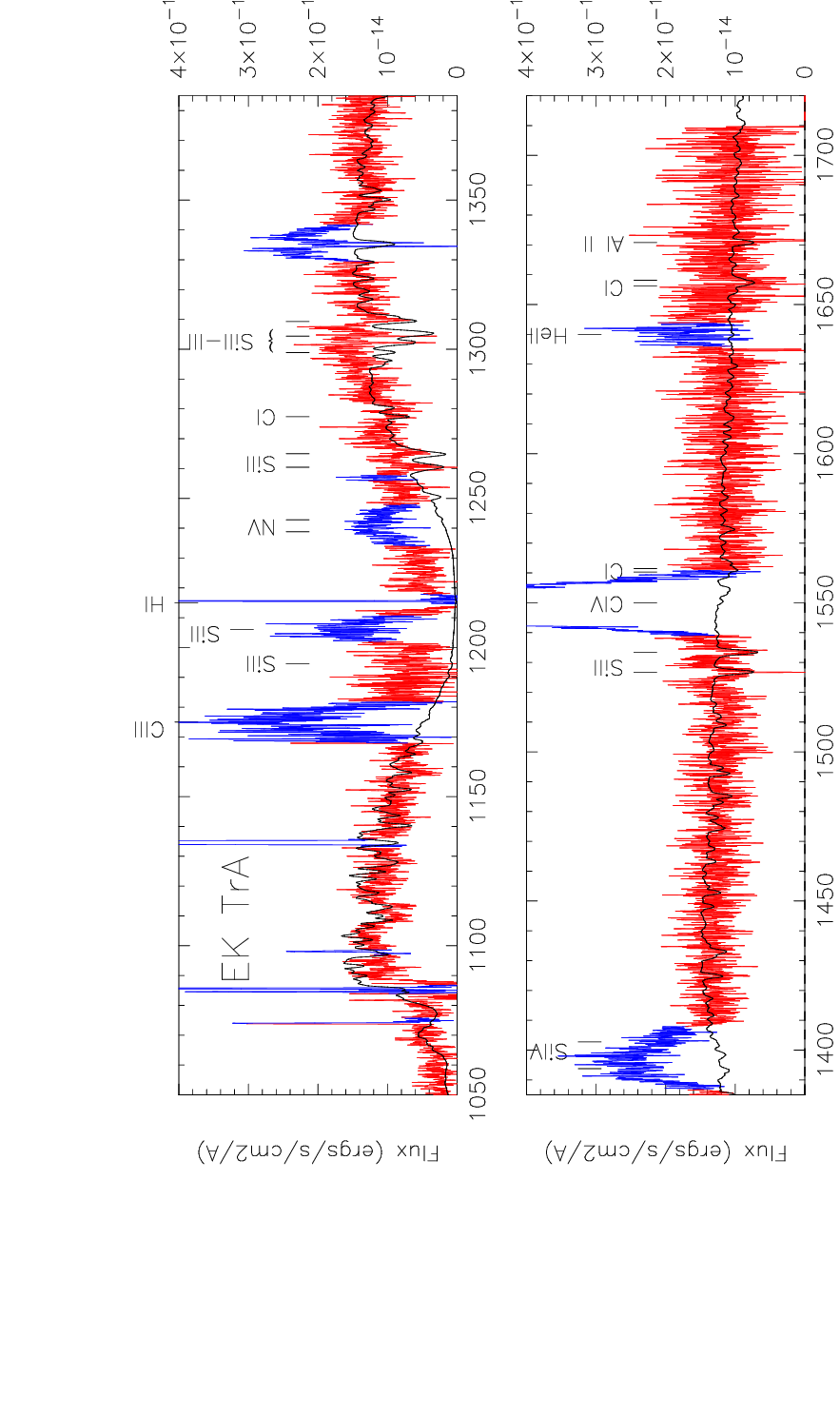}                                     
\end{figure}

\clearpage 

\begin{figure}
\plotone{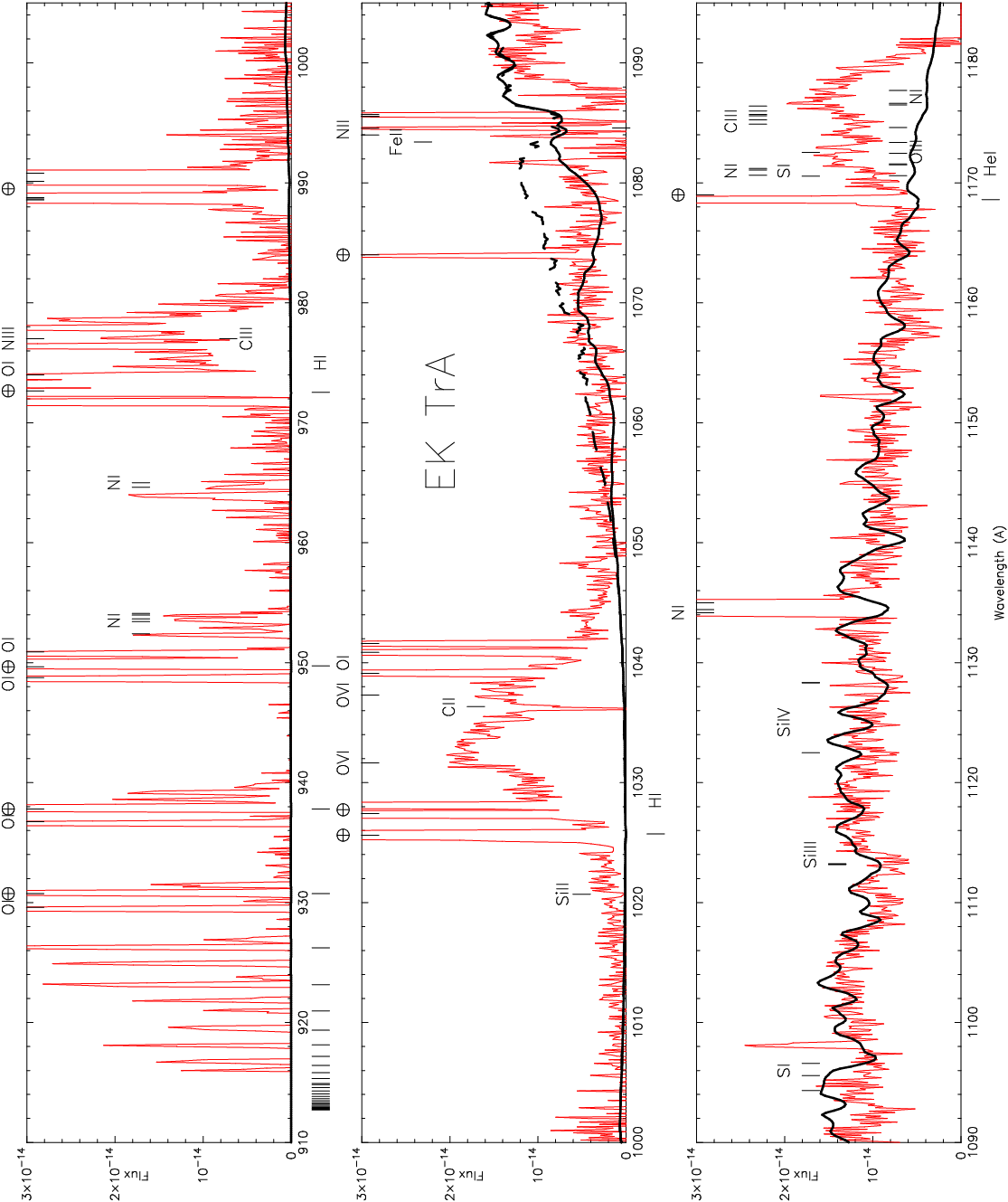}                                     
\end{figure}

\clearpage 

\begin{figure}
\plotone{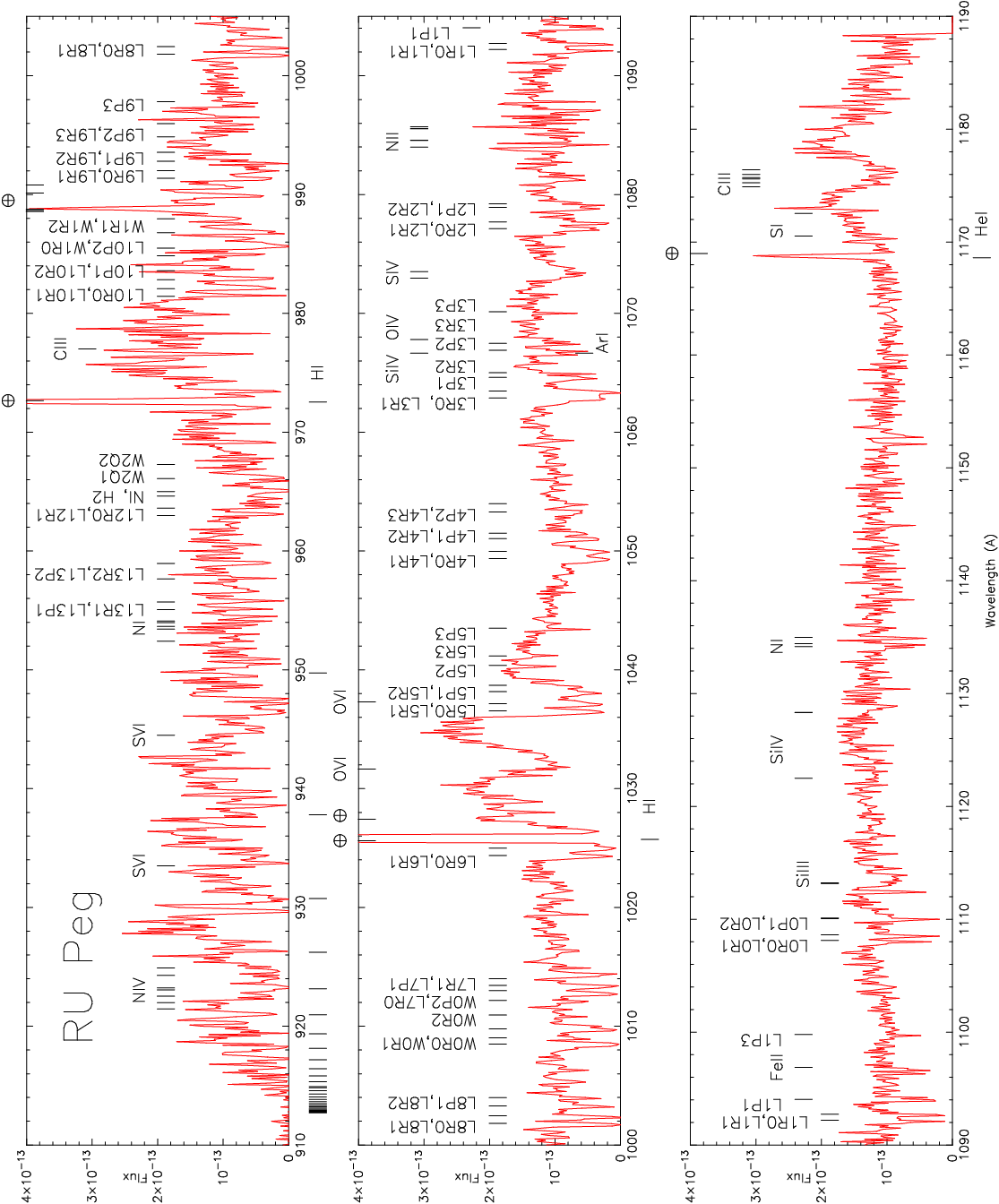}                                     
\end{figure} 

\clearpage 

\begin{figure}
\plotone{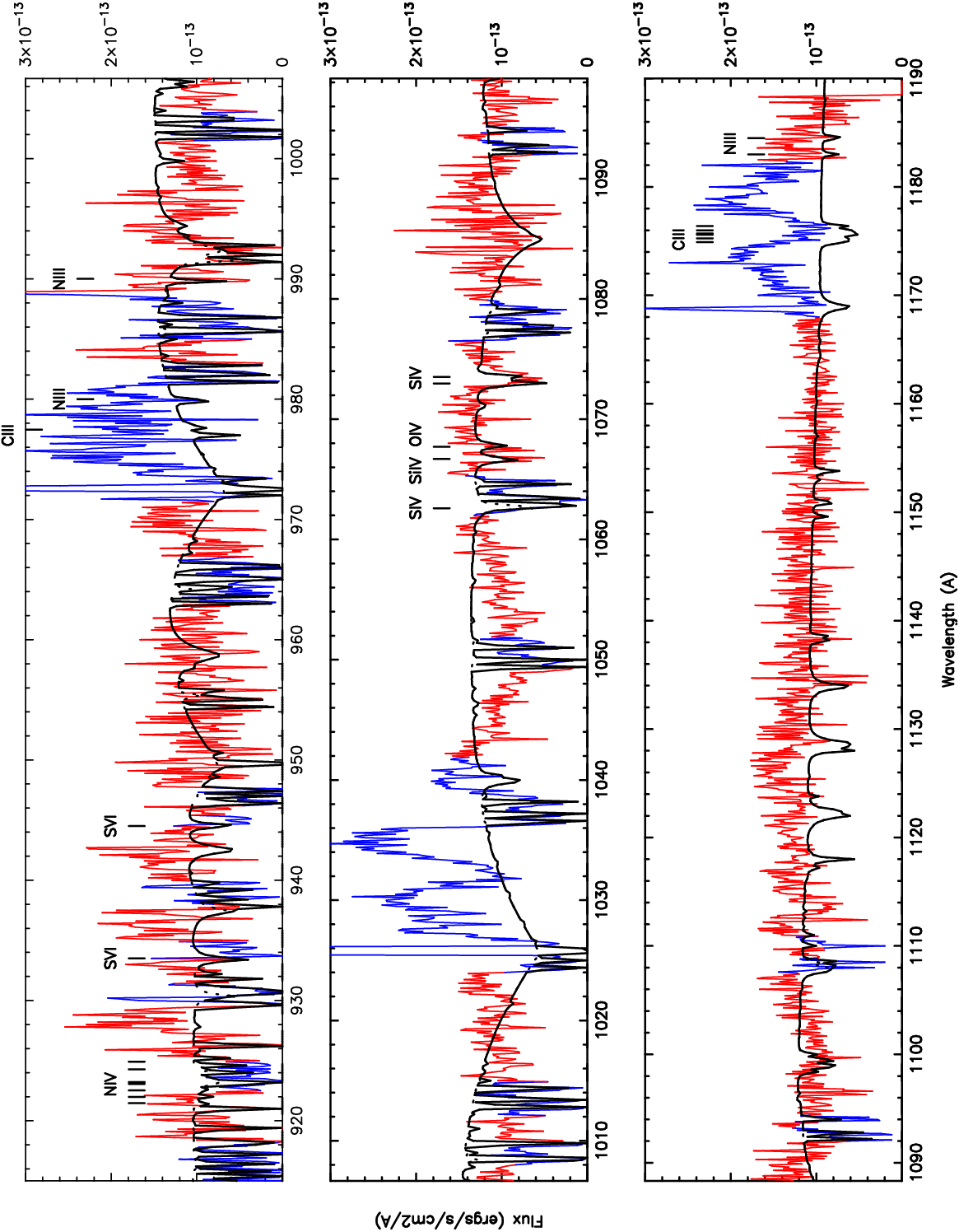}                                     
\end{figure} 

\clearpage 

\begin{figure}
\plotone{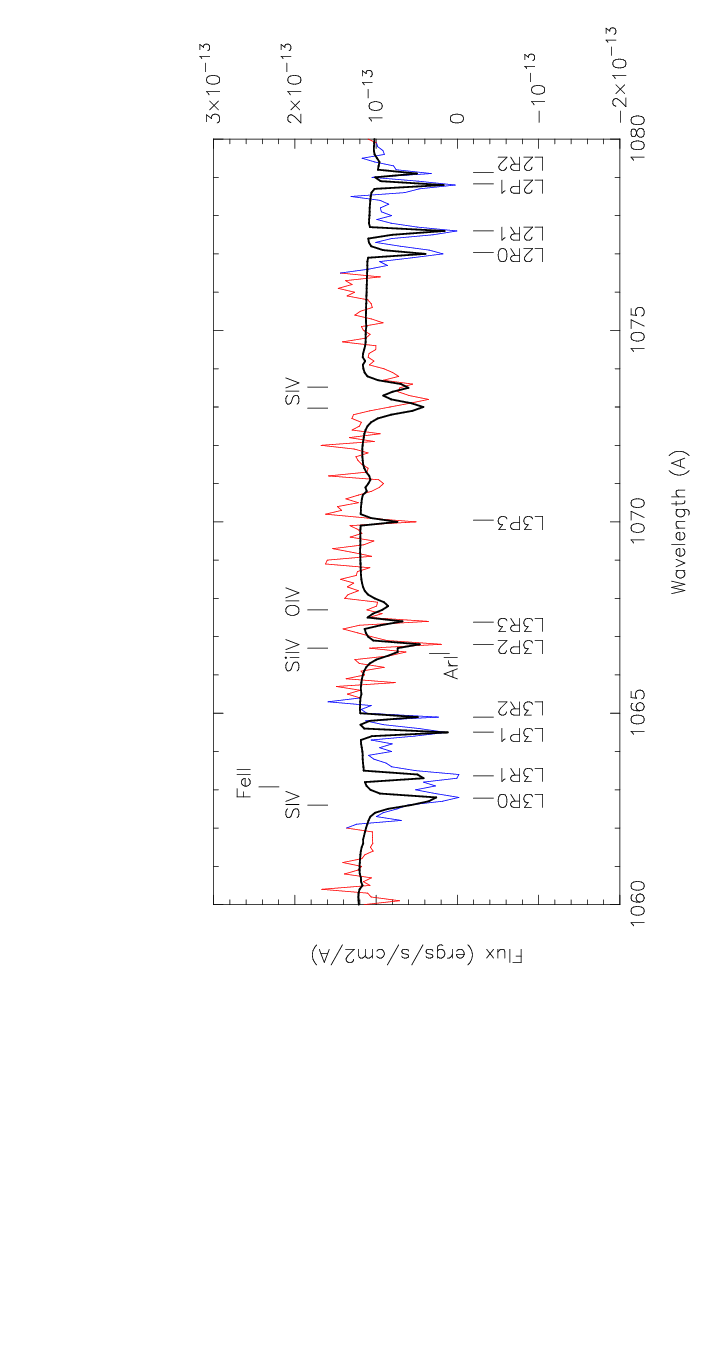}                                     
\end{figure}

\end{document}